**Nanodosimetric investigation of the track structure of therapeutic carbon ion radiation. Part 2: Detailed radiation transport and track structure simulation.**

Miriam Schwarze*, Gerhard Hilgers, Hans Rabus

Physikalisch-Technische Bundesanstalt, Braunschweig and Berlin, Germany

* corresponding author, miriam.schwarze@ptb.de

**Abstract**

*Objective:* Previously reported nanodosimetric measurements of therapeutic-energy carbon ions penetrating simulated tissue have produced results that are incompatible with the predicted mean energy of the carbon ions in the nanodosimeter and previous experiments with lower energy monoenergetic beams. The purpose of this study is to explore the origin of these discrepancies.

*Approach:* Detailed simulations using the Geant4 toolkit were performed to investigate the radiation field in the nanodosimeter and provide input data for track structure simulations, which were performed with a developed version of the PTra code.

*Main results:* The Geant4 simulations show that with the narrow-beam geometry employed in the experiment, only a small fraction of the carbon ions traverse the nanodosimeter and their mean energy is between 12 % and 30 % lower than the targeted values. Only about one-third or less of these carbon ions hit the trigger detector. The track structure simulations indicate that the observed enhanced ionization cluster sizes are mainly due to coincidences with events in which carbon ions miss the trigger detector. In addition, the discrepancies observed for high absorber thicknesses of carbon ions traversing the target volume could be explained by assuming an increase in thickness or interaction cross-sections in the order of 1 %.

*Significance:* The results show that even with strong collimation of the radiation field, future nanodosimetric measurements of clinical carbon ion beams will require large trigger detectors to register all events with carbon ions traversing the nanodosimeter. Energy loss calculations of the primary beam in the absorbers are insufficient and should be replaced by detailed simulations when planning such experiments. Uncertainties of the interaction cross-sections in simulation codes may shift the Bragg peak position.

## 1. Introduction

Conventional methods for radiation treatment planning consider the tissue type and energy spectrum of the particles to calculate the macroscopic average dose to tumor or organ tissue. However, the effectiveness of ionizing radiation is related to the spatial distribution of energy depositions in microscopic biomolecular targets, such as the DNA molecule or parts of it [1]. Nanodosimetry therefore investigates the stochastic nature of energy depositions in nanometer-sized volumes. As concepts such as the average energy imparted per ionization are not applicable for such small target dimensions [2], experimental nanodosimetry focuses on the ionization component of the charged particle track structure and determines frequency distributions of the number of ionizations in nanometric targets. With the recent advances towards the implementation of nanodosimetric parameters of the radiation field obtained by simulations into treatment planning [3–7], the provision of experimental reference data of nanodosimetric track structure parameters of ion beams under clinical conditions (i.e. for the mixed radiation field present within the human body) has become a necessity.

In the frame of nanodosimetry, the microscopic structure of the track is described by the distribution of the relative frequency of ionization cluster size (ICS), $P_\nu(Q,d)$. The ICS $\nu$ is the number of ionizations by a primary particle of radiation quality $Q$ and all secondary particles in a target located at a distance $d$ from the track of the primary particle. The $i$-th moment of the ICS distribution is labeled $M_i(Q,d)$.

In the first part of the paper [8], nanodosimetric experiments with clinical carbon ions behind absorbers representing human tissue were reported. Ionization cluster size distributions (ICSDs) were measured at different depths along a pristine Bragg peak. Larger ICSs than expected were obtained from the carbon ion energy values calculated with the SRIM (**S**topping power and **R**ange of **I**ons in **M**atter) code [9,10] or earlier measurements of monoenergetic carbon ion beams in the frame of the BioQuaRT (biologically weighted quantities in radiotherapy) project [11]. Furthermore, the conditional $M_1$ increased for high $d$ values, which was not recognizable in the BioQuaRT measurements. To investigate the effects of secondary particles on the nanodosimeter signal, further measurements with different combinations of primary particle energy and absorber thickness were performed. The absorber thickness and primary energy were combined to ensure that the carbon ions in the target volume of the nanodosimeter had identical energy, according to calculations with the SRIM code. Due to the identical energy of the carbon ions in the target volume, the same ICSDs were expected for carbon ions traversing the target. Instead, an almost constant shift to larger $M_1$ values of several tens of percent was observed with increasing primary energy and absorber thickness. This increase seemed too large to be explained by the different secondary particle backgrounds alone.

In order to gain a better understanding of the observed effects, simulations of the interactions of the radiation field with the experimental setup are presented in this part of the paper. The simulation was divided into two steps. First, a condensed history simulation of the particles in the entire measurement setup was performed using Geant4 [12–14]. The information obtained serves as input data for the second simulation, a track structure simulation, using the PTra (PTB track structure) code [15–18], which simulates the interaction of the particles in the target volume of the nanodosimeter in detail. The simulation studies were conducted following the analysis of the experimental data.

## 2. Materials and Methods

### *2.1 Condensed-history simulation of the radiation field in the nanodosimeter*

The setup used for the Geant4 condensed-history simulations is shown in Figure 1. The carbon ion beam has a Gaussian shape with a standard deviation of 4.25 mm (which corresponds to a full width at half maximum of 10 mm). The beam first passes through a cuboid PMMA absorber with a side length of 30 cm and varying thickness according to the thickness used in the experiment. A cuboid PMMA collimator with an aperture of $(2 \times 10)$ mm$^2$, a side length of 30 cm, and a thickness of 10 cm is placed downstream. The beam enters the nanodosimeter through a 5 mm thick cuboid entrance window made of A150 plastic with a side length of 20 cm. Behind the entrance window, the nanodosimeter is represented by a gas volume comprising propane at a pressure of 1.2 mbar. The gas volume ends at the end of the defined geometry. The capacitor plates used in the nanodosimeter for target gas ion extraction are omitted in the simulation. Instead, a plane 13 cm behind the entrance window is inserted at the location of the extraction hole in the lower plate, which defines the target volume in the experiments. Particles are scored when passing this plane within a circle of 16.5 mm radius around the beam axis (which is the radial distance of the extraction orifice from the beam axis in the experiment). The area of this circle will be referred to as the target volume aperture (TVA) and the plane as the target volume plane (TVP). Two strip detectors with a cross-section of $(2 \times 10)$ mm$^2$ are placed in the gas volume as a 0.3 mm slab of silicon and a 1 mm slab of polyvinyl chloride.

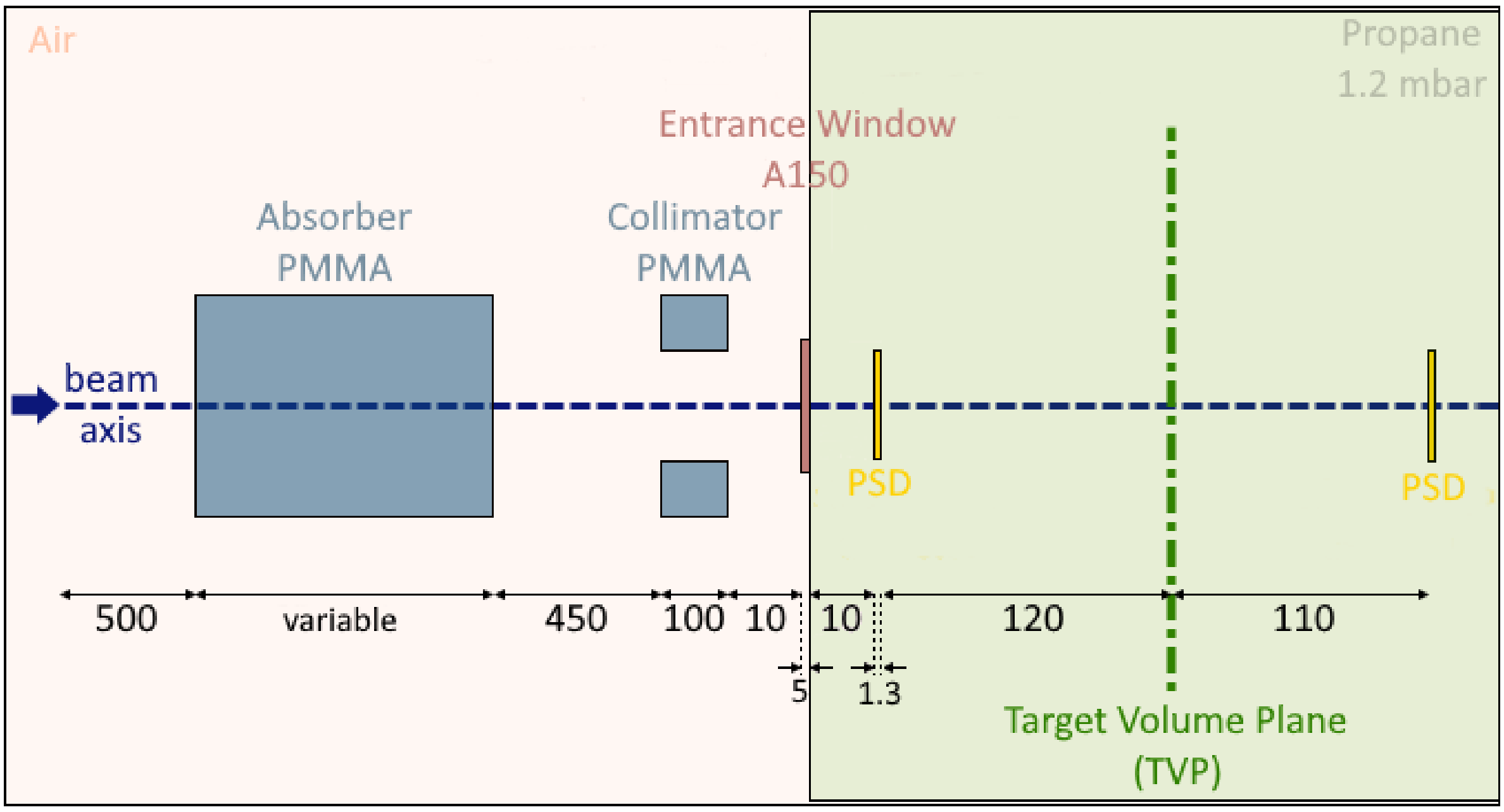

Figure 1. Cross-sectional view of the simplified setup used for the Geant4 simulations (not to scale). The lengths along the beam direction are given in millimeters. The sides of the rectangular collimator aperture and silicon strip detectors are the same and amount to 10 mm in one direction perpendicular to the beam and 2 mm in the other direction. The outer dimensions of the absorber and collimator along these directions are 300 mm, and the outer dimension of the propane gas volume are 3000 mm. The target volume aperture (not shown) is located within a radius of 16.5 mm around the beam axis in the TVP.

The simulations were executed on the PTB high-performance cluster with Geant4 version 11.0.2, each involving $10^8$ histories. Electromagnetic processes were included by the *G4EmStandardPhysics_option4* – constructor, decay processes by the *G4DecayPhysics* – constructor, and hadronic processes by the models *G4HadronElasticPhysicsHP, G4IonQMDPhysics, G4HadronPhysicsShielding, G4EmExtraPhysics, G4StoppingPhysics, G4RadioactiveDecayPhysics and G4NeutronTrackingCut.* The range cut, which specifies a lower limit of the range for the generation of new secondary particles, was set to 1 mm for all particles.

The position, direction of motion, energy, particle type, and event identification number (EventID) were scored for each particle passing the TVA. For all interaction processes in the silicon detectors, the position, energy, particle type, and EventID were recorded.

### *2.2 Track structure simulation in the nanodosimeter*

Since conducting a track structure simulation throughout the entire measurement setup was prohibitively time-consuming, a two-step approach was taken, combining a condensed history simulation of the radiation transport up to the nanodosimeter and a detailed track structure simulation within its interaction volume.

To obtain the particle fluence of the radiation field, the condensed history simulation with Geant4 was repeated using the setup shown in Figure 1, although the range cut for electrons was reduced to 0.02 mm (corresponding to a production energy cut for electrons of about 10 eV in propane gas). All particles were registered when they crossed the TVP. The positions before and after the crossing were written to a list file along with the particle type and energy at the start point of the step crossing the TVP.

The track structure simulations were performed using the PTra code[1] that had been developed to simulate nanodosimetric measurements using nitrogen, propane, or water vapor as operating gases [32–35]. To process the output files from the Geant4 simulation, the PTra code was developed starting from version PTra_c3h8_20170608 for the Ion Counter nanodosimeter operated with propane. All subroutines and functions related to cross-section evaluation, random sampling, the simulation of interaction processes, and the general structure of the main code for particle tracking remained unchanged. The development comprised a change of the primary particle source to data read from a phase space file and implementing a different approach to tallying.

The approach is illustrated in Figure 2. The Geant4 simulations gave the positions of the particles before and after passing the TVP (Figure 2(a)). Electrons were started at the last point before crossing the TVP. The trajectories of carbon ions and other heavy-charged particles were back-projected to the plane at the rear of the first PSD (Figure 2(b)), from where the simulation of their tracks was started (Figure 2(c)). The ionizations produced by the tracks of all particles of the same event were then scored with the detection efficiency map centered on the different positions of a detector array (Figure 2(d)), which had an increment in impact parameter of 0.25 mm and a spacing along the beam direction of 1.5 mm.

Additional modifications were made to further improve the scoring statistics. First, all tracks were scored a second time after inverting all vertical positions, thus exploiting the mirror symmetry of the simulation setup. Second, instead of considering the efficiency map via a Russian roulette approach to determine whether an ionization is detected, the probabilities of the number of detected ions were directly estimated for each event. For this purpose, the detection probabilities of the efficiency map were successively used to construct the probability distribution of the number of ionizations for the event (based on the locations of the points of ionizing interactions with respect to the efficiency map). The corresponding procedure is illustrated in Supplementary Figure S1. Finally, an average across all events was calculated.

Photons and neutrons were ignored in the track structure simulations, which is justified by their low interaction probability in a dilute gas. Sparsely occurring leptons and neutrinos were also ignored, while all sparsely occurring heavy charged particles (such as, e.g., beryllium or oxygen ions) were included.

---

[1] Using PTra instead of Geant4 was motivated by the fact that the Geant4-DNA toolkit [19–22] was originally developed for track structure simulations in liquid water. While cross-sections of DNA constituents and nitrogen gas measured in the BioQuaRT project [23–27] are now available in the Geant4-DNA code [28,29], the implementation of the measured propane cross-sections [30,31] is still in progress.

The following three classes of events were scored separately: (i) events with a carbon ion having a trajectory intersecting the second PSD and depositing energy in it triggering the data acquisition in the experiment (triggered events), (ii) events with a carbon ion in the interaction volume of the nanodosimeter following a trajectory missing the second PSD (outlier events), and (iii) events containing only electrons or heavy charged particles other than carbon (no-carbon events). The simulations of outlier and no-carbon events were undertaken to provide data for estimating the background contribution from events arriving in coincidence with events in which a carbon ion triggered data collection in the nanodosimeter. Their union is referred to as non-triggered events.

For triggered events, the plane defined by the direction of the carbon ion and its projection on the horizontal *x*-*z*-plane was used as the reference, and targets were placed evenly spaced along a direction perpendicular to this reference plane (Figure 2(d)) and along the intersection of the reference plane with the horizontal plane. In this way, ICSDs at different impact parameters could be scored simultaneously. This approach is justified by the assumption that the pattern of ionizations in a track should not change when the track is displaced laterally or subjected to a small rotation about an (almost) perpendicular axis. Furthermore, the variation of the ICSDs along the beam could be assessed, and averaging over different targets could be performed at the same impact parameter if this variation was as small as expected (to improve the scoring statistics).

The mean ICS $M_1$ was determined according to

$$M_1 = \sum_{\nu=0}^{\infty} \nu \, P_\nu \quad , \tag{1}$$

where $P_\nu$ is the relative frequency of targets scoring $\nu$ ionizations.

For non-triggered events, the common vertical symmetry plane of the two PSDs was used as the reference plane. In these cases, eventually only the targets with a 3 mm offset from the axis of the initial carbon ion beam were considered when determining the background contribution. For scoring ionizations in non-triggered events, the efficiency map corresponding to a semi-infinite time window was used.

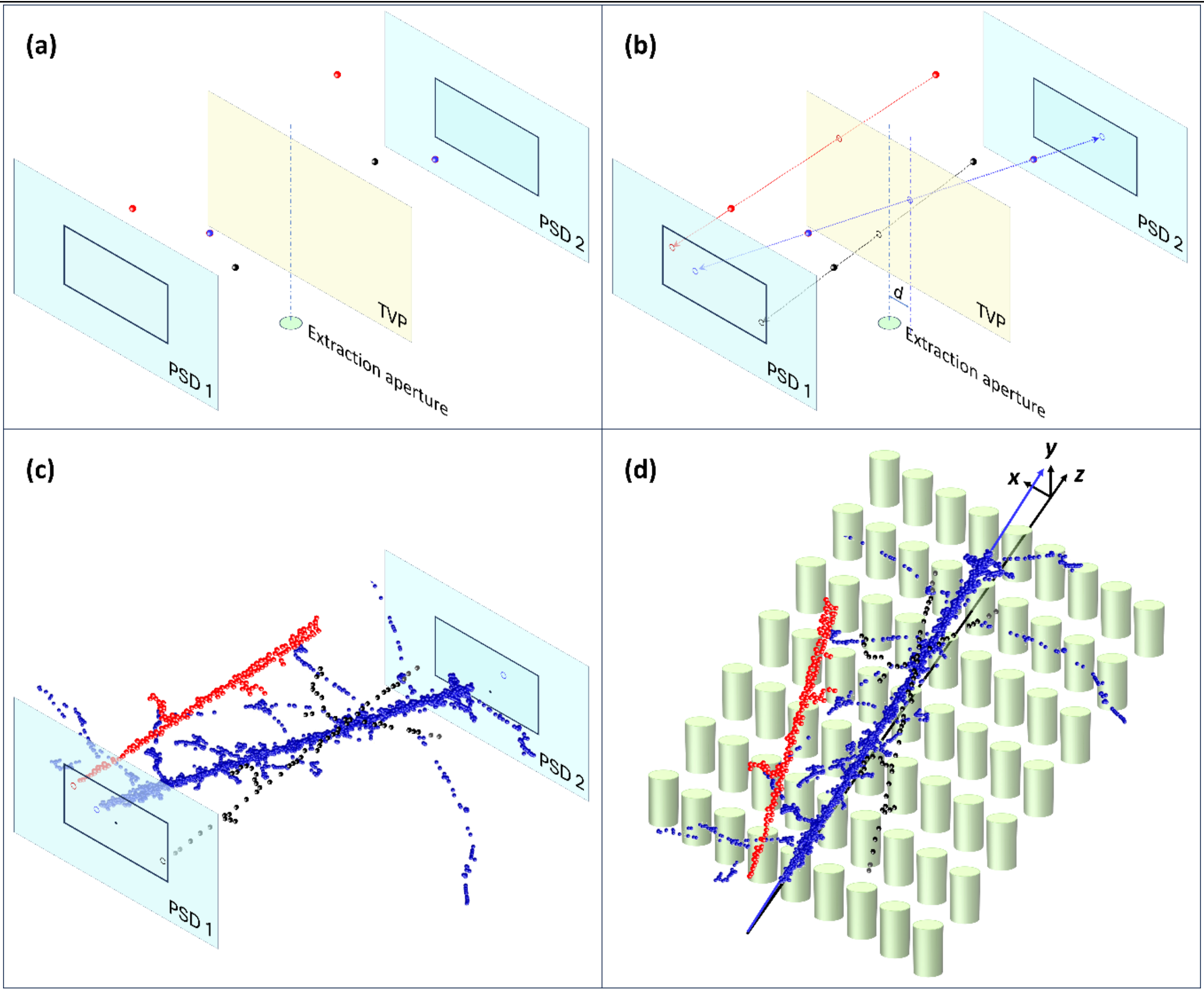

Figure 2. Schematic illustration of the approach used in the track structure simulations using PTra. (a) Interaction points from the Geant4 simulation of a carbon ion (blue dots) and two secondary heavy-charged particles (red, black) before and after passing the TVP. (b) The particle trajectories are projected to the rear of the first silicon strip detector (PSD1, open circles). (c) The particle tracks are started in the PTra simulation from the rear of PSD1 (open circles) with the kinetic energy at the first interaction points in (a) along the direction of the projection lines in (b). Ionizations produced in the tracks of the carbon ions (blue dots) and two secondary particles (red and black dots), by these particles or their secondary electrons, are recorded. (d) The array of target volumes is aligned such that the trajectory of the carbon ion (blue line) is in its mirror-symmetry plane (*y*-*z*-plane), and the ionization clusters are scored in all targets.

## 3. Results

### *3.1 Energy distributions of carbon ions in the nanodosimeter*

In the first part of the paper [8], indications of incorrect energy calculations using SRIM were discovered. To verify the energy calculations by SRIM, the corresponding energy distributions are compared to energy distributions calculated by Geant4, which allows creating a more detailed geometry and may have more accurate or at least more comprehensive cross-section data. The comparison is presented in Figure 3, with the Geant4 simulation results depicted as energy distributions in the respective bottom plots in the form of solid lines. All carbon isotopes that traverse the TVA are included. The energy spectra calculated with SRIM are shown in the respective upper plots. It is evident that the energies calculated by Geant4 are smaller than the values calculated by SRIM.

Figure 3(a) shows the case of the measurements with constant primary energy and varying absorber thickness. A decrease is evident in both the energies calculated by SRIM and Geant4, which is expected due to the increasing energy loss of carbon ions in the absorber. However, the energy distributions from Geant4 are consistently shifted to lower energy values, in the worst case by almost a factor of two.

For the measurements with combined variation of primary carbon ion energy and absorber thickness, presented in Figure 3(b), the energies determined by Geant4 show a decrease in energy in the target volume with increasing primary energy and absorber thickness. This decrease can be compared with the recorded pulse height spectra of the second PSD from the experiment, which were depicted in Figure 11 in the first part of the paper [8]. Due to the small thickness of the PSD, the carbon ions are not fully stopped in the detector, as they deposit only a small fraction of their kinetic energy. Therefore, the measured peak channel number is proportional to the stopping power, which increases with decreasing energy of the particles in the PSD. The peak was found to be shifted towards larger channels with increasing initial kinetic energy and PMMA absorber thickness, indicating an increasing energy loss in the detector and thus a decreasing carbon ion energy. This finding is more consistent with the Geant4-calculated energies, both indicating a decrease in the energy of the carbon ions with increasing primary energy and absorber thickness.

The carbon ion energies calculated by Geant4 support the suspicion that the energy values calculated with SRIM were not representing the actual values in the experiments. The Geant4 simulation results indicate that an energy smaller than the SRIM calculated energy in the target volume was achieved in both experiments. The lower energy leads to an increasing LET and thus an increasing $M_1$. The planned measurement strategy of generating a constant carbon ion energy in the nanodosimeter with different secondary particle fields was therefore not achieved in these measurements. For this reason, the remainder of the paper does not further discuss these measurements and focuses on those with constant primary carbon ion energy.

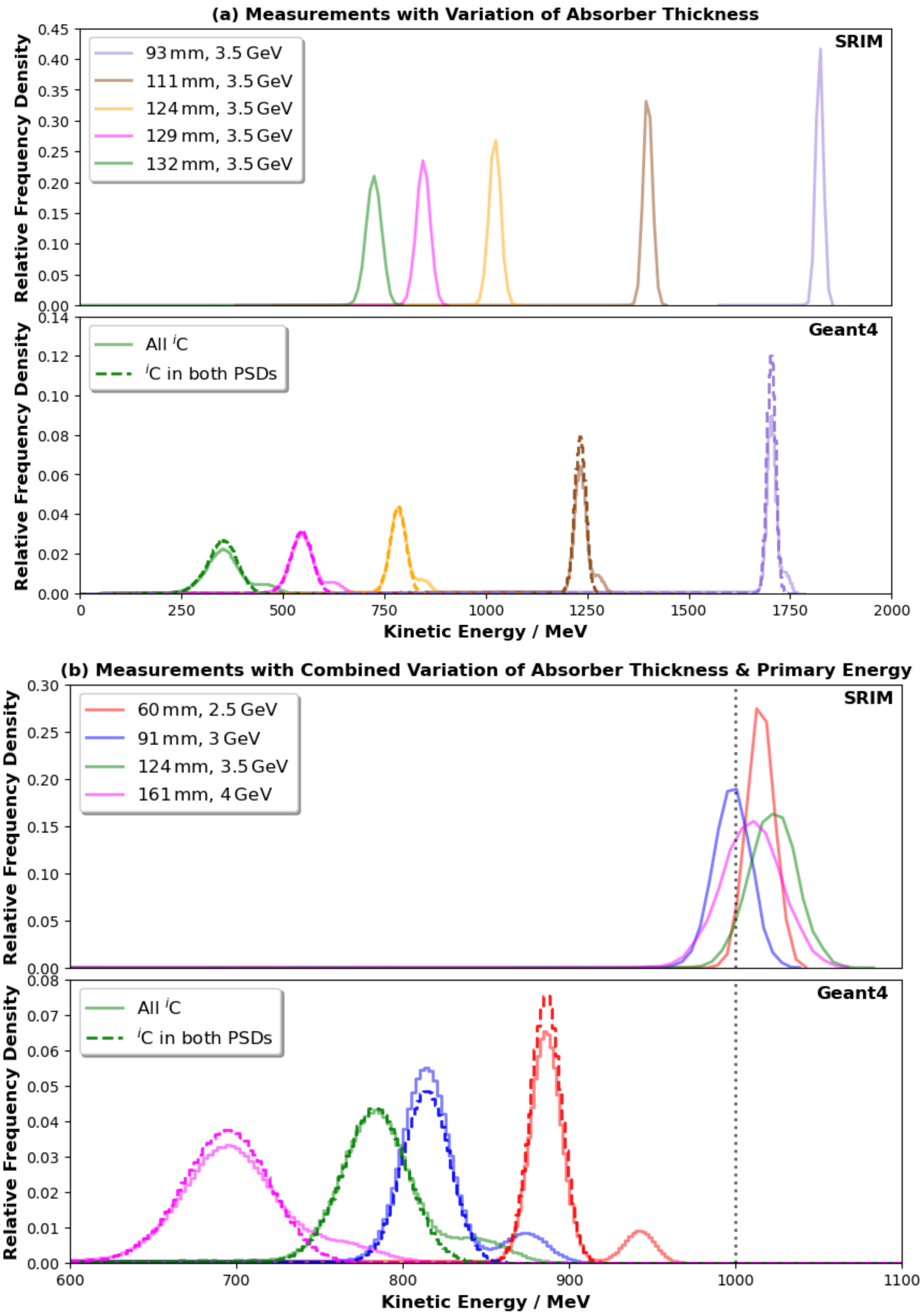

Figure 3. Comparison of energy spectra calculated by SRIM and Geant4 for the measurements with constant primary carbon ion energy and variation of absorber thickness (a) and the measurements with combined variation of primary carbon ion energy and absorber thickness (b). The respective upper plot represents the SRIM calculated energy spectra. The respective bottom plot shows the energy spectra calculated with Geant4 of all carbon isotopes passing the TVA. The solid lines depict the spectra for all events, and the dashed lines relate to events with a carbon ion interacting in both PSDs.

In Figure 3, a small higher-energy satellite peak is visible next to each of the main peaks in the Geant4 results. Carbon ions contributing to these peaks did not interact with the PSDs. This becomes apparent when considering only the energy spectra of carbon ions that traversed both PSDs, as depicted by the dashed lines in Figure 3. Since the second PSD serves as a trigger detector, only events in which a carbon ion interacts in the second PSD were considered in the experiment. All other events may result in enhanced $M_1$ values if they occur in coincidence with such triggered events and produce additional ionizations. The frequency of such events is examined in more detail in the following sections.

### *3.2 Spatial distribution of carbon ions*

To explore the contribution of outlier events, Figure 4 shows the spatial distributions of carbon ions in different planes perpendicular to the initial beam direction. Figure 4(a) and (b) show the distribution of carbon ion positions at the rear of the first PSD, and Figure 4(c) and (d) at the front of the second PSD. The data were obtained by extrapolating the connecting line of the last interaction before and the first after passing the TVA, as illustrated in Figure 2(b). The gray shaded areas represent the detector dimensions. It is evident that the second PSD, which serves as trigger detector, is overfilled by the carbon ion beam, and a large proportion of carbon ions miss this detector for all five absorber thicknesses.

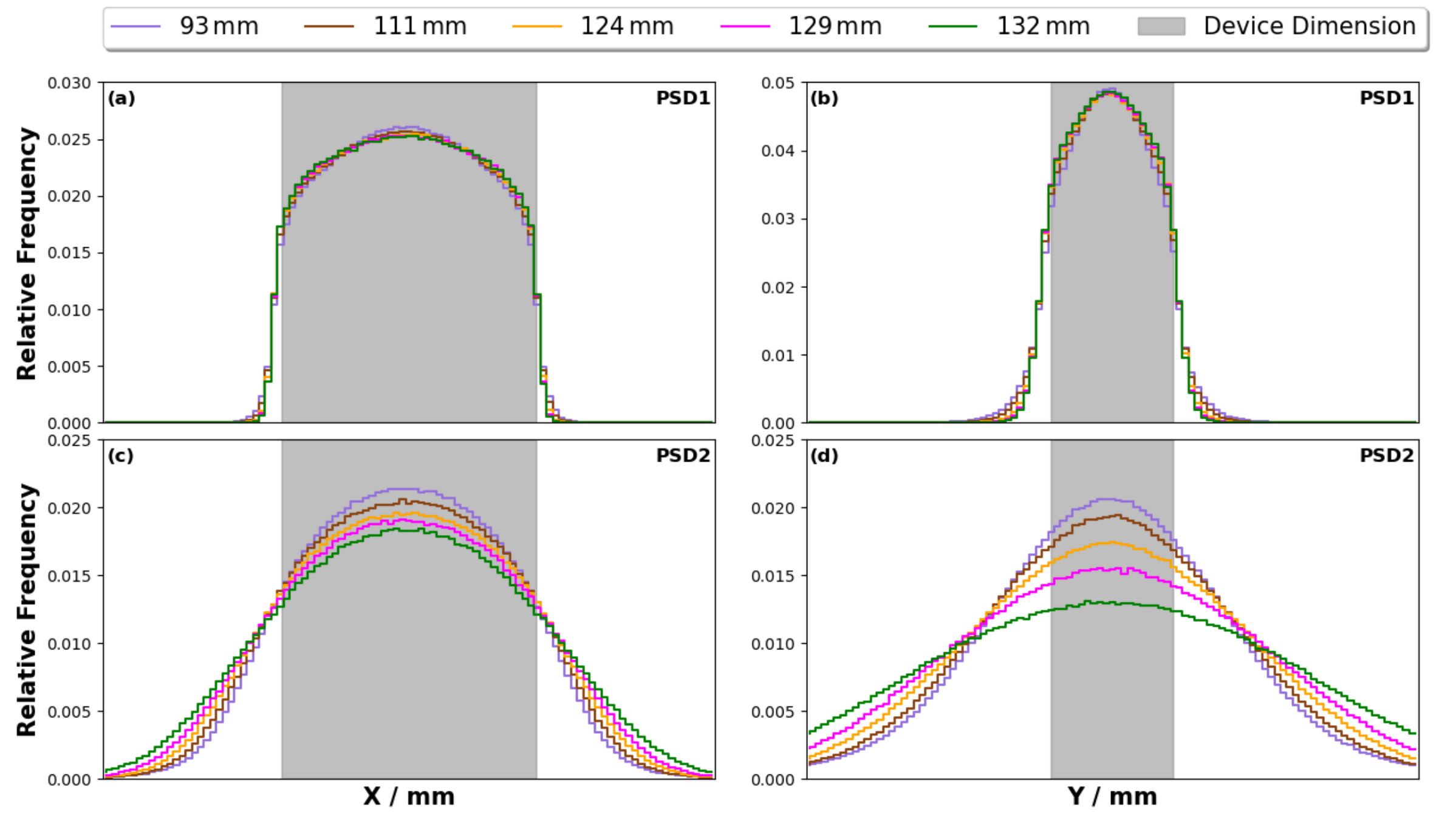

Figure 4. Distribution of the projected carbon ion positions at the PSDs and the collimator. (a) and (b): position at the rear of the first PSD along its (a) larger and (b) smaller dimensions. (c) and (d): at the front of the second PSD along its (c) larger and (d) smaller dimensions.

The proportion of events involving carbon ions that traverse both detectors is listed in the first column of Table 1. Carbon ions missing the first PSD also always miss the second one. The proportion of carbon ions entering the nanodosimeter through the collimator aperture is less than 6 % (third column of Table 1). Both proportions decrease with increasing absorber thickness, i.e. with decreasing primary carbon ion energy, as seen in the first two columns of Table 1. However, the proportion of events in which any charged particle traverses the TVA is between about 6 % and about 13 %, as can be seen in the last column of Table 1.

Non-triggered events relate only to the absence of a carbon ion traversing the second PSD. To investigate the impact of other heavy charged particles traversing the trigger detector, the distributions of energy imparted per event in the second PSD is shown in Supplementary Figure S3 for carbon ions and the most abundant secondary charged particles. Boron and lithium have been included as they were considered as possible candidates of secondary heavy charged particles potentially imparting sufficient energy to trigger a measurement. In the experimental study of the charged particle field associated with $^{12}C$ beams in a water phantom by Haettner et al. [36], boron ions were found to make a substantial contribution to the secondary particle field. As can be seen in Supplementary Figure S3, the energy imparted by secondary heavy charged particles traversing the second PSD is generally by an order of magnitude smaller than that by carbon ions. A noticeable contribution from boron ions traversing this PSD is only observable for the thickest absorber. However, the corresponding distribution of energy imparted is at values below 25 % of the peak of the energy distribution from carbon ions. Therefore, such events would be discriminated in the experiment by the window discriminator covering the primary carbon ion peak in the energy spectrum, and it can be concluded that all triggered events in the experiments were due to carbon ions traversing the second PSD.

Table 1. Overview of the proportions of different event categories obtained from the Geant4 simulations for the different absorber thicknesses. The event numbers are normalized to the total number of simulated events in each case. First column: Events with a carbon ion hitting both PSDs (triggered event). Second column: Events with a carbon ion missing the second PSD (outlier events). Third columns: Events with a carbon ion crossing the TVA (sum of columns 1 and 2). Fourth column: Events without a carbon ion but other charged particles crossing the TVA (no-carbon events). Fifth column: Events with any charged particles crossing the TVA (sum of columns 3 and 4).

| **Absorber thickness / mm** | **(1) Triggered events** | **(2) Outlier events** | **(3) Events with carbon ion in TVA** | **(4) No-carbon events** | **(5) Events with any charged particle in TVA** |
|---|---|---|---|---|---|
| 93 | $1.75 \times 10^{-2}$ | $3.81 \times 10^{-2}$ | $5.56 \times 10^{-2}$ | $6.98 \times 10^{-2}$ | $12.52 \times 10^{-2}$ |
| 111 | $1.10 \times 10^{-2}$ | $2.78 \times 10^{-2}$ | $3.88 \times 10^{-2}$ | $4.91 \times 10^{-2}$ | $8.79 \times 10^{-2}$ |
| 124 | $0.67 \times 10^{-2}$ | $2.05 \times 10^{-2}$ | $2.72 \times 10^{-2}$ | $4.06 \times 10^{-2}$ | $6.78 \times 10^{-2}$ |
| 129 | $0.49\times 10^{-2}$ | $1.18 \times 10^{-2}$ | $2.27\times 10^{-2}$ | $3.84 \times 10^{-2}$ | $6.11 \times 10^{-2}$ |
| 132 | $0.34 \times 10^{-2}$ | $1.64 \times 10^{-2}$ | $1.98 \times 10^{-2}$ | $3.70 \times 10^{-2}$ | $5.68 \times 10^{-2}$ |

### *3.3 Secondary particle field*

In the experiments presented in the first part of the paper [8], increased $M_1$ values were observed compared to the reference data. One reason for this was suspected to be an increased number of ionizations by high LET secondary particles. This hypothesis is further investigated here by examining the stopping power of carbon ions and the most frequent secondary particles crossing the TVA (the frequency of the different particles is depicted in Supplementary Figure S2). The stopping power values were calculated using the Bethe-Bloch equation [37], utilizing the energy values obtained from the Geant4 simulation. As observed in the first part of the paper [8], a comparable energy dependence of stopping power and $M_1$ can be assumed. Thus, the stopping power can be expected to be proportional to the measured signal.

The mean values of the stopping powers are shown in Figure 5. In Figure 5(a), triggered events are shown, and in Figure 5(b) non-triggered events. The total stopping power of all particles increases continuously with increasing absorber thickness. The increased stopping power leads to an increased number of ionizations and thus a higher average signal in the nanodosimeter. For the triggered events, the overwhelmingly dominant contribution to the total stopping power is due to carbon ions. A minimal contribution of protons can be observed for all configurations, as well as an even smaller contribution of deuterons in configurations with high absorber thicknesses. This suggests that secondary particles only contribute marginally to the signal of the triggered events.

For the non-triggered events, heavy charged secondary particles also contribute to the total stopping power. However, the highest contribution is due to carbon ions, being almost one order of magnitude higher than the next highest contribution from alpha particles.

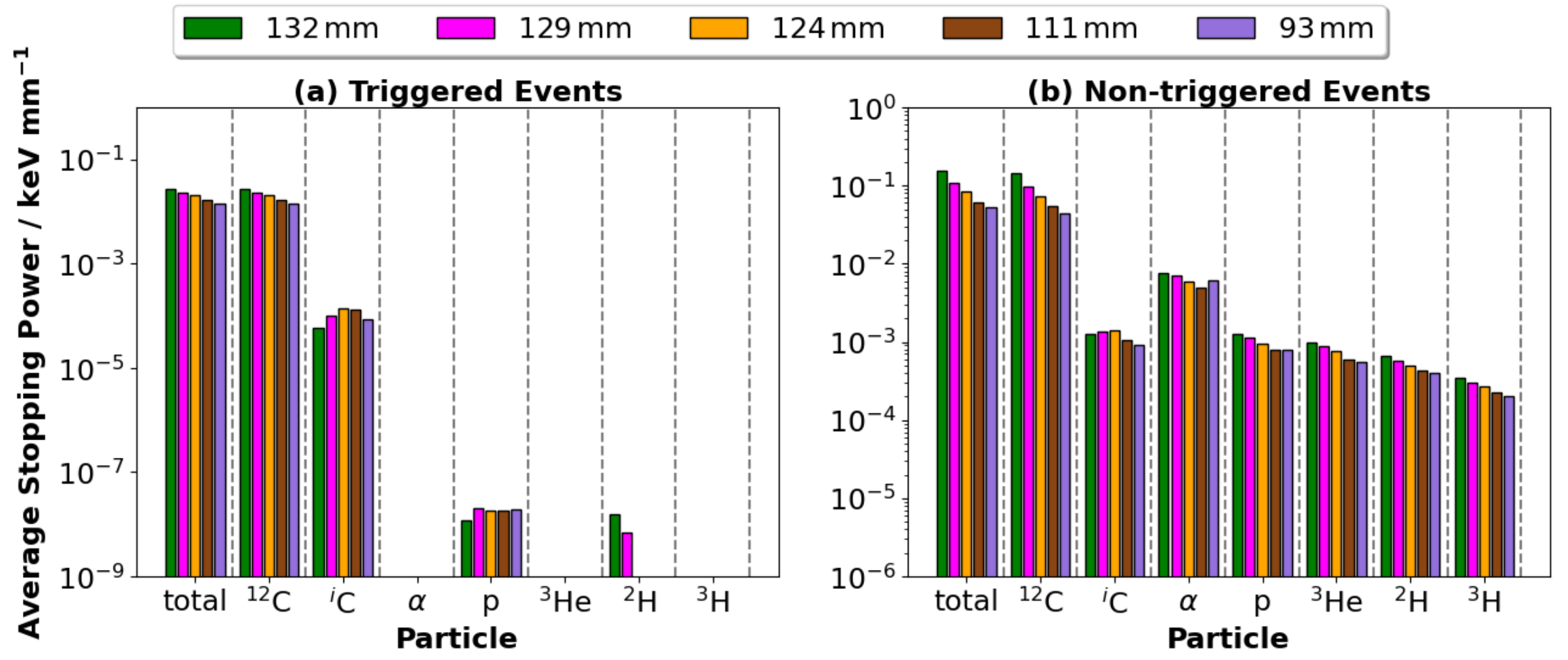

Figure 5: Average total stopping power of the most abundant heavy charged particles traversing the TVA and the contributions from carbon ions and the most frequent secondary particles passing the TVA. (a): events with a carbon ion hitting PSD2; (b): events without hit in PSD2. Both histograms are normalized to the total number of carbon ions crossing the TVA.

### *3.4 Track structure simulations: Estimate of the background contributions*

In section 3.2, it was observed that a significant proportion of the carbon ions miss the second PSD, which serves as trigger detector. Events in which the carbon ion is absorbed in the components before the nanodosimeter and only some of its secondary charged particles traverse the nanodosimeter occur even more frequently than events with carbon ions missing the trigger detector (Table 1).

The potential background contributions to $M_1$ from these events was estimated based on the PTra track structure simulation data for non-triggered events. Here, only the results from the targets located at $z$ positions between -5 mm and +5 mm relative to the extraction aperture position and at $y$ positions of ± 3 mm (horizontal offset of the extraction aperture in the experiments) were averaged. In contrast to triggered events, for which the $M_1(d)$ values scored in the target array shown in Figure 2(d) were found to be independent of $z$, the $M_1$ values scored for outlier events can be seen to decrease by about 10 % along the 140 mm path length shown in Supplementary Figure S. This variation is attributed to two factors. First, a proportion of these carbon ions has trajectories at a larger angle with respect to the beam axis than carbon ions hitting both PSDs. Therefore, along a segment of the path through the interaction volume, the carbon ion is found at positions outside the range of extraction efficiency. Second, the carbon ions that miss only the second PSD have undergone interactions in the first PSD. Among these is a fraction of carbon ions with energies in the lower wing of the energy distributions with higher ionization cross-sections. Owing to this variation with the z position of the targets, only targets in the vicinity of the position of the extraction aperture in the measurements were considered (indicated by the yellow symbols in Supplementary Figure S(b)).

If a carbon ion missing the second PSD arrives in coincidence with a carbon ion registered in the second PSD, the directions of motion of the two carbon ions can be assumed to be uncorrelated. The same can be assumed for secondary particles from a carbon ion absorbed in the setup before reaching the nanodosimeter. Therefore, the same contribution of outlier events or no-carbon events can be assumed for all impact parameters of carbon ions hitting the second PSD and triggering a measurement.

The probability of coincidence of a carbon ion triggering a measurement and a background event was estimated as follows. In the experiments, the typical count rate of events in the trigger detector was below 1000 $s^{-1}$ with a maximum count rate of up to 2000 $s^{-1}$ [8]. The drift time window employed for collecting the propane gas ions had a width of 56 µs. Therefore, the probability of a coincidence between a carbon ion hitting both PSDs and one missing the second PSD can be roughly estimated at 1/20. The same estimate applies to the probability of coincidence between a carbon ion hitting both PSDs and the charged particles of an event with no carbon ion in the nanodosimeter.

### 3.4.1 Background contribution from outlier events.

The results of the corresponding track structure simulations, shown in the first column of Table 2, demonstrate that the $M_1$ per outlier event grows by almost a factor of three when the PMMA absorber thickness varies between 93 mm and 132 mm (for the initial kinetic carbon ion energy of 3.5 GeV).

Table 2. Contributions to $M_1$ from events in which a carbon ion traverses the interaction volume without hitting the second PSD (outlier events) for a carbon ion beam of 3.5 GeV initial kinetic energy and different PMMA absorber thickness. First column: Average $M_1$ per outlier event. Second column: Ratio of the number of outlier events to the number of triggered events. Third column: Estimated frequency of coincidences between outlier and triggered events. Fourth column: Resulting background contribution to the measured $M_1$.

| **Absorber thickness / mm** | **(1) Average $M_1$ per outlier event** | **(2) Ratio outlier to triggered events** | **(3) Coincidence Frequency** | **(4) Estimated $M_1$ background** |
|---|---|---|---|---|
| 93 | 0.316 ± 0.002 | 2.38 ± 0.08 | 0.13 ± 0.04 | 0.04 ± 0.01 |
| 111 | 0.377 ± 0.002 | 2.74 ± 0.09 | 0.15 ± 0.04 | 0.06 ± 0.02 |
| 124 | 0.520 ± 0.004 | 3.30 ± 0.11 | 0.18 ± 0.05 | 0.10 ± 0.03 |
| 129 | 0.687 ± 0.005 | 3.96 ± 0.14 | 0.22 ± 0.06 | 0.15 ± 0.04 |
| 132 | 0.987 ± 0.009 | 5.15 ± 0.18 | 0.29 ± 0.08 | 0.29 ± 0.08 |

The second column of Table 2 shows the ratio of the number of outlier events to the number of triggered events. This ratio shows an increase by more than a factor of two between the smallest and largest absorber thicknesses. The values correspond to the ratios of the numbers in the second column of Table 1 to those in the first column corrected for the reduction of the denominator resulting from an estimated potential misalignment by $\pm$0.2 mm of the PSDs relative to the collimator aperture.

Considering the 56 µs time window for data acquisition, a mean rate of 1000 triggered events per second, and the ratio of outlier events to triggered events in column 2, the expected frequencies of coincidences between outlier and triggered events shown in the third column of Table 2 are obtained. The uncertainty estimate in this column includes the uncertainty of the rate of triggered events as a further component. A symmetric triangular distribution was assumed for the event rate peaking at 1000 $s^{-1}$ and extending up to the highest rate of 2000 $s^{-1}$ encountered in the experiments [38].

The last column in Table 2 shows the resulting estimate of the background due to coincidently arriving carbon ions that miss the trigger detector. The values are obtained by multiplying the average $M_1$ per outlier event with the frequency of coincidences.

### 3.4.2 Background contribution from no-carbon events in the nanodosimeter.

The background contribution of events in which the carbon ion is absorbed in the components before the nanodosimeter but some of its secondary charged particles traverse the nanodosimeter was estimated using the same analysis as for the carbon ions in Table 2, with the results shown in Table 3.

Table 3. Estimated contribution to $M_1$ from events where only electrons or heavy-charged particles other than carbon ions traverse the TVA. First column: Average $M_1$ per event without carbon ion entering the nanodosimeter (no-carbon events). Second column: Ratio of the number of no-carbon events to the number of triggered events. Third column: Mean coincidence frequency of a no-carbon event and a triggered event. Fourth column: Resulting background contribution to the measured $M_1$.

| **Absorber thickness / mm** | **(1) Average $M_1$ per no-carbon event** | **(2) Ratio of no-carbon to triggered events** | **(3) Coincidence Frequency** | **(4) Estimated $M_1$ background** |
|---|---|---|---|---|
| 93 | $(1.64 \pm 0.04)\times10^{-2}$ | 4.4 ± 0.2 | 0.24 ± 0.07 | $(0.40 \pm 0.12)\times10^{-2}$ |
| 111 | $(1.45 \pm 0.03)\times10^{-2}$ | 4.9 ± 0.2 | 0.27 ± 0.08 | $(0.40 \pm 0.12)\times10^{-2}$ |
| 124 | $(1.45 \pm 0.03)\times10^{-2}$ | 6.6 ± 0.2 | 0.37 ± 0.11 | $(0.54 \pm 0.16)\times10^{-2}$ |
| 129 | $(1.48 \pm 0.03)\times10^{-2}$ | 8.7 ± 0.3 | 0.49 ± 0.14 | $(0.7 \pm 0.2)\times10^{-2}$ |
| 132 | $(1.52 \pm 0.04)\times10^{-2}$ | 11.9 ± 0.4 | 0.67 ± 0.19 | $(1.0 \pm 0.3)\times10^{-2}$ |

The resulting $M_1$ values in the first column of Table 3 are one to two orders of magnitude smaller than those seen in the first column of Table 2 because only a small fraction of events without carbon ions contains charged particle trajectories intersecting the target volume. The ratio of no-carbon events to triggered events (second column of Table 3) is about a factor of two higher than for outlier events and increases more strongly with increasing absorber thickness. Consequently, the number of coincidences per triggered event is also about a factor of two higher than for outlier events and rises to about 0.66 for the thickest absorber. Owing to the low number of ionizations on average (column 1 of Table 3), the resulting background contribution (last column of Table 3) is between a factor of sixteen (at 93 mm absorber thickness) and 4.5 (132 mm) smaller than the contribution from outlier events.

However, it should be noted, that in this case the same extraction efficiency map as for the carbon ion event simulations was used because the efficiency maps had only been determined for the corresponding drift time window. Therefore, this contribution and its uncertainty may be underestimated since it is not considered that the charged particles other than carbon ions have larger deviations from the initial carbon beam axis (Supplementary Figure S). As they traverse regions where the extraction efficiency may differ from the extraction efficiency that applies to the secondary ions produced by primary carbon ions, the uncertainties of the background contribution may be assumed to be larger than shown in Table 3.

The sum of the estimated background values in the last rows of Table 2 and Table 3 is between $0.05 \pm 0.01$ at 93 mm and about $0.30 \pm 0.08$ at 132 mm absorber thickness.

### *3.5 Track structure simulations: Dependence of ICS on impact parameter*

From the track structure simulations for triggered events, ICSDs were obtained for each target in the target array shown in Figure 2(d) and then averaged over the targets at the same impact parameter at *z* positions between 92 mm and 227 mm behind the entrance window. No noticeable variation of $M_1$ scored in the targets placed at different positions along the beam direction was observed (cf. Supplementary Figure S(a)).

Figure 6 shows a comparison of the measured data (blue circles) and the simulation results (red diamonds) for the dependence of $M_1$ on impact parameter $d$ for the five absorber thicknesses employed with the 3.5 GeV carbon ion beam. Ionizations produced by electrons and heavy charged particles from these events are included in the simulation results.

It can be observed that in all cases there is a strong discrepancy between the measured values and the simulation results at impact parameters exceeding 2 mm. In this range of impact parameters, the simulation produces a continuous decrease of $M_1$, whereas almost constant values can be seen for the measured data.

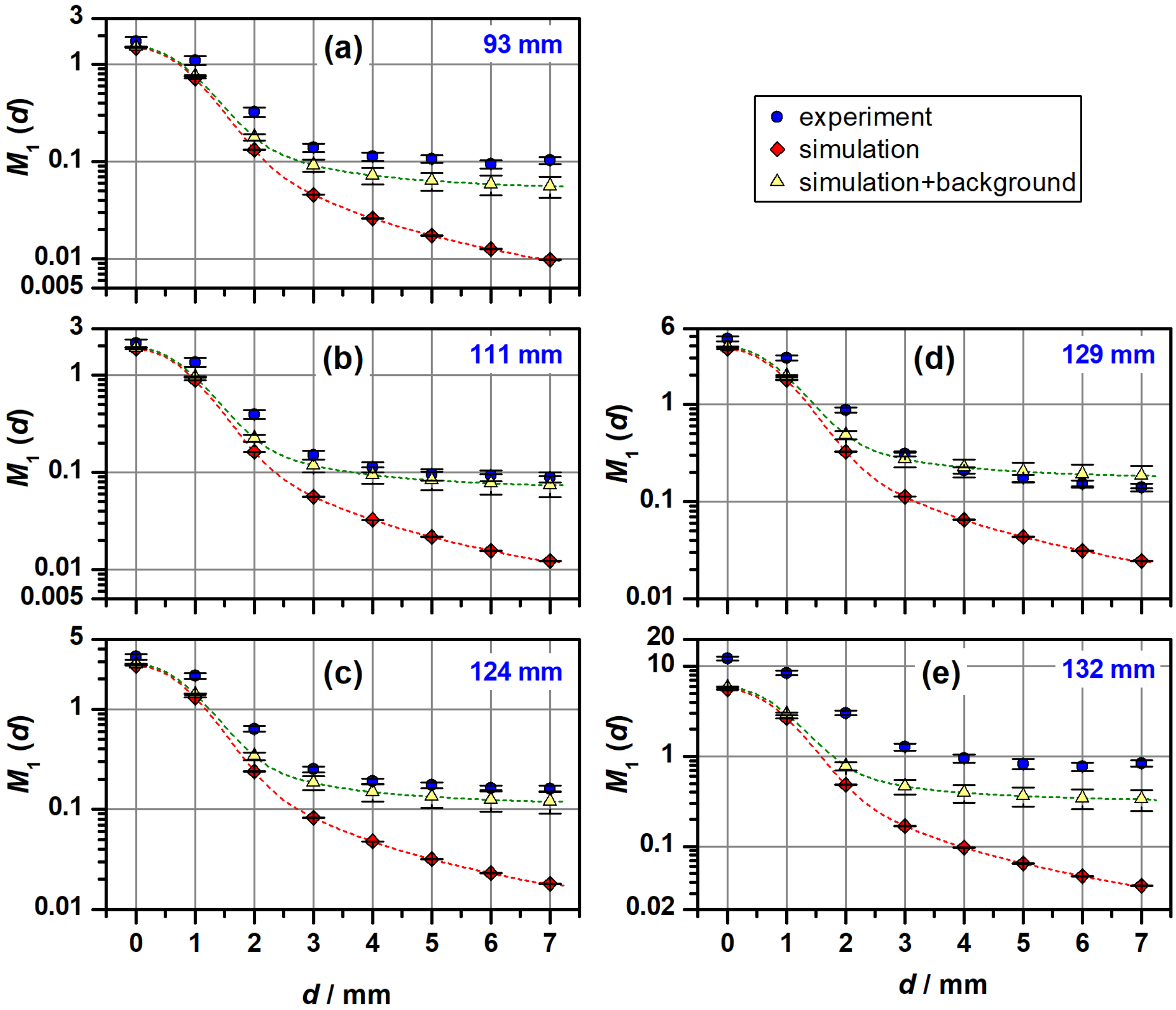

Figure 6: Dependence of $M_1$ on impact parameter $d$ for different thickness of PMMA absorber in front of the nanodosimeter and an initial kinetic energy of carbon ions of 3.5 GeV: Results of the track structure simulations with PTra (red diamonds) compared to the experimental results (blue circles). The red lines represent the averaged simulation data over intervals of 1 mm in impact parameter. The yellow triangles represent the background-corrected simulation data.

In addition, there is also a discrepancy between measurements and simulations at impact parameters below 2 mm, which strongly increases with increasing thickness of the absorber.

At 132 mm absorber thickness, the experimental data for $M_1(d = 0 \text{ mm})$ are more than a factor of two larger than the values obtained by the simulations.

As demonstrated by the data plotted as yellow triangles in Figure 6, for which the estimated background was added, the discrepancy at large impact parameters can be explained in most cases by the constant background estimated in the previous section.

In the case of 132 mm PMMA absorber thickness shown in Figure 6(e), the background correction leads to a similar relative variation with the impact parameter, while an overall discrepancy by a factor exceeding two remains. For all absorber thicknesses, the background correction does not significantly change the discrepancies at impact parameters smaller than 2 mm.

### 3.6 Track structure simulations: Variation of ICS with absorber thickness

Figure 7 shows a comparison of simulation and measurement results for the variation of $M_1$ at impact parameter $d = 0$ mm with the thicknesses of the absorbers used in the experiment. The blue circles represent the measured data, while the red diamonds are the results of PTra simulations considering only events in which a carbon ion hits the second PSD. The yellow triangles represent the simulation data after applying the background correction estimated in section 3.4.

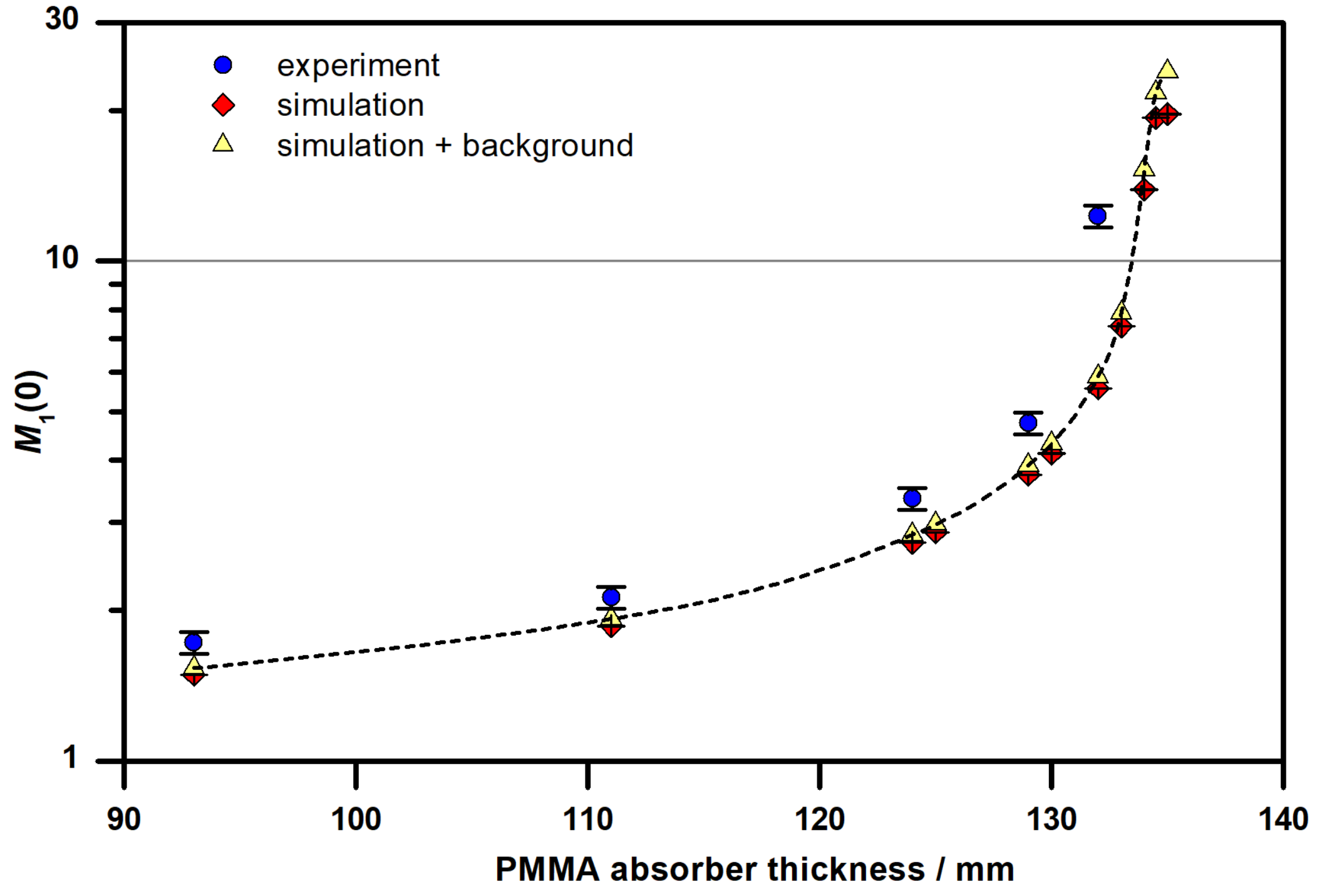

Figure 7: Variation of the mean ICS for the central passage of the carbon ion through the target volume, $M_1(0)$, with absorber thickness. Blue circles: measurement; red diamonds: PTra simulations without background correction; yellow triangles: PTra simulation including the background corrections given in Table 2 and Table 3. The dashed line is a guide to the eye through the background-corrected simulation data.

It can be seen from Figure 7 that after applying the background correction, the discrepancies between the measured data and the simulation results persist. This is to be expected since the magnitude of the corrections shown in Table 2 and Table 3 are much smaller than the differences between simulation and experiment in Figure 6 at zero impact parameter. Considering the results of simulations at additional thickness values that are also shown in Figure 7, it appears that most of the discrepancy between experiment and simulation can be attributed to a shift of the abscissa in Figure 7 in the order of 1 to 2 mm or about 0.8 % to 1.6 % of the thickness of the absorber assumed in the simulations.

Figure 8 shows a comparison of the measured and simulated ICSDs for $d = 0$ mm for the five PMMA absorber thicknesses used in the experiments at 3.5 GeV initial kinetic carbon ion energy. The blue circles are the experimental data, and the red diamonds represent the simulation results from PTra for the corresponding absorber thickness. It can be seen that for the two smallest thicknesses (Figures 8(a) and (b)), the ICSDs are very similar except for a significantly higher probability at ICS zero in the simulations and a resulting reduction at ICS different from zero, which accumulates to the underestimation of the measured $M_1(0)$ seen in Figure 7.

For 124 mm PMMA absorber thickness (Figure 8(c)), the simulated ICSD is higher than the measured one for ICS of zero and 1, and below the experimental values for ICS above 2. Increasing the thickness of the absorber in the simulation slightly by 1 mm (green triangles) reduces the discrepancies, although overall the simulated ICSD remains significantly different from the measured one.

Figure 8(d) shows the case of 129 mm PMMA absorber thickness where the simulated ICS frequencies are higher up to an ICS of 3. In this case, the simulation for 130 mm instead of 129 mm absorber thickness gives an ICSD that matches the measured distribution much better than the simulation for 129 mm.

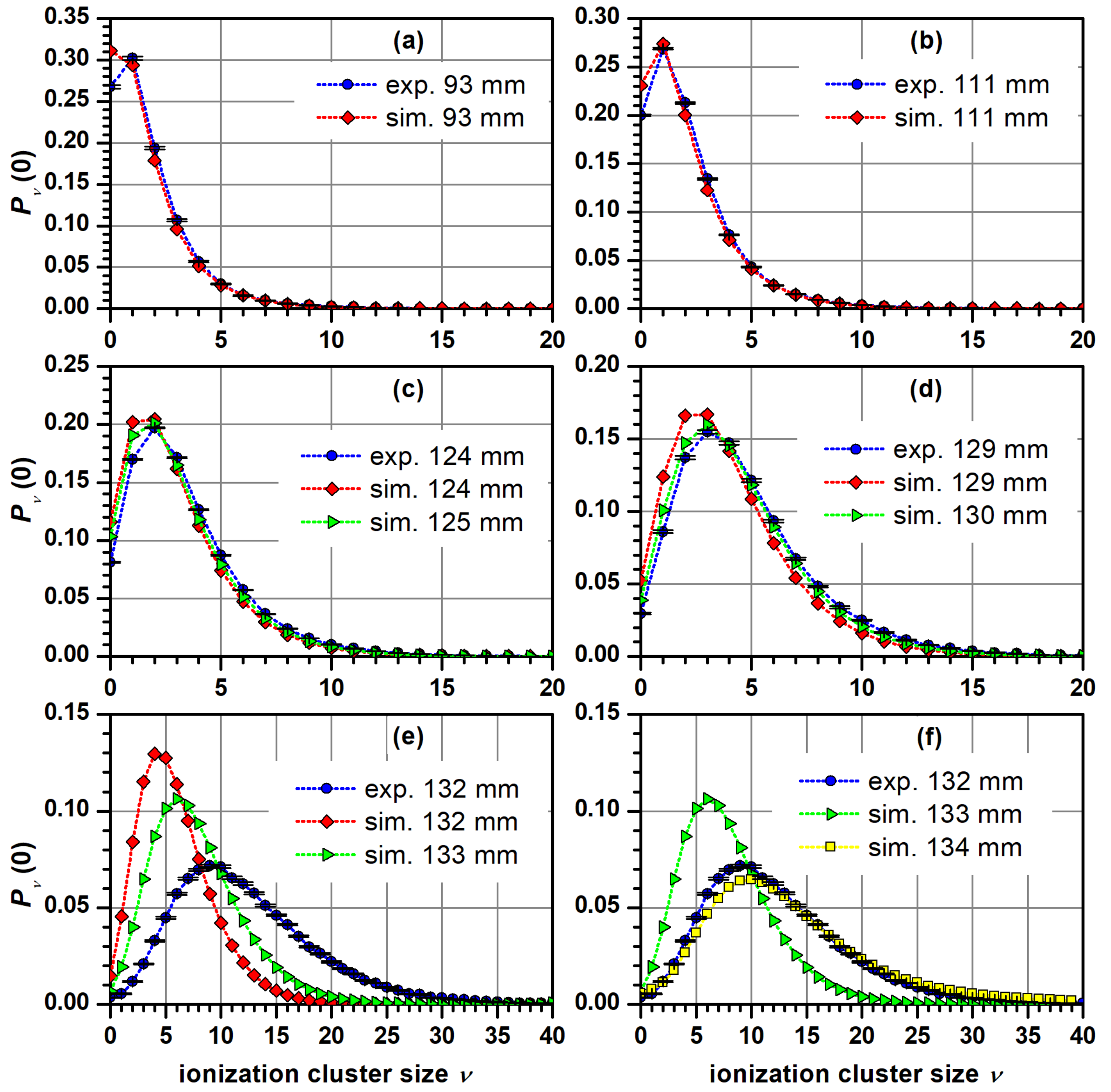

Figure 8: Comparison of the measured ICSDs (blue circles) and the simulations for the same PMMA absorber thickness nominally used in the experiments (red diamonds). The simulation results in (c) to (f) marked by green triangles are for a PMMA absorber thickness increased by 1 mm; the yellow square in (f) correspond to a 2 mm increase in PMMA thickness.

For the case of the largest absorber thickness, the ICSD from the simulation for the 132 mm thickness used in the experiments no longer resembles the experimental one (Figure 8(e)). In this case, increasing the thickness in the simulation by 1 mm still gives an ICSD far from the experimental one. However, as can be seen from Figure 8(f), increasing the thickness by 2 mm gives a distribution very close to the measured one. This ICSD appears to be shifted slightly to higher ICS values and suggests that a thickness close to but smaller than 134 mm would have produced a perfect match.

The background in the ICSDs due to coincidences is not considered in this analysis and would shift the distributions to slightly higher ionization cluster sizes.

## 4. Discussion

The purpose of this simulation study was to understand the peculiarities in the experimental results reported in the first paper [8]. The dependence of ICS on impact parameter and the relative energy dependence of the $M_1(0)$ were at variance with experiments from the BioQuaRT project [35] and the relative energy dependence of stopping power values calculated by SRIM. The ionization clusters from the new measurements appeared to be significantly increased. In addition, the measured peak channel spectra from the second PSD indicated that the energy of the carbon ions in the nanodosimeter may have significantly deviated from the values calculated with SRIM when the experiment was designed.

### *4.1 Energy of the carbon ions in the nanodosimeter*

The Geant4 simulation results presented in Figure 3 confirm the suspicion that the carbon ion energy in the nanodosimeter is overestimated by the SRIM calculations used during the experiment planning phase. Other studies have also documented deficiencies in energy calculation with SRIM [39,40]. The discrepancy between the energy distributions calculated by Geant4 and SRIM increases with increasing absorber thickness. This leads to the conclusion that in the measurements with combined variation of absorber thickness and primary carbon ion energy, no constant carbon ion energy was achieved in the nanodosimeter, but rather a decreasing energy with increasing absorber thickness. Consequently, the intended measurement strategy to compare different secondary particle fields with the same interaction behavior of the carbon ions in the nanodosimeters was not achieved in the experiment.

In the first paper [8], discrepancies were found between the relative energy dependence of the measured $M_1(0)$ values for the measurements with constant primary carbon ion energy, experiments from the BioQuaRT project [35] as well as stopping power values calculated by SRIM. In addition, the measured values appeared to be significantly increased. These deviations are mainly caused by the incorrect energy values calculated by SRIM. This is illustrated in Figure 9, which shows the measured $M_1(0)$ values as a function of carbon ion energy in the nanodosimeter calculated by SRIM (filled circles) and Geant4 (open circles) in comparison with the BioQuaRT results (blue) and the mass stopping power of carbon ions in propane (red line). Using the Geant4 calculated energy values shifts the data points to lower energy values and removes most of the discrepancies except for the thickest PMMA absorbers. Plotting the respective $M_1(0)$ value with the mean energy from the Geant4 simulations for 130 mm and 134 mm absorber thickness (cyan-filled circles) brings these datapoints in good agreement with the trend of the BioQuaRT and stopping power data.

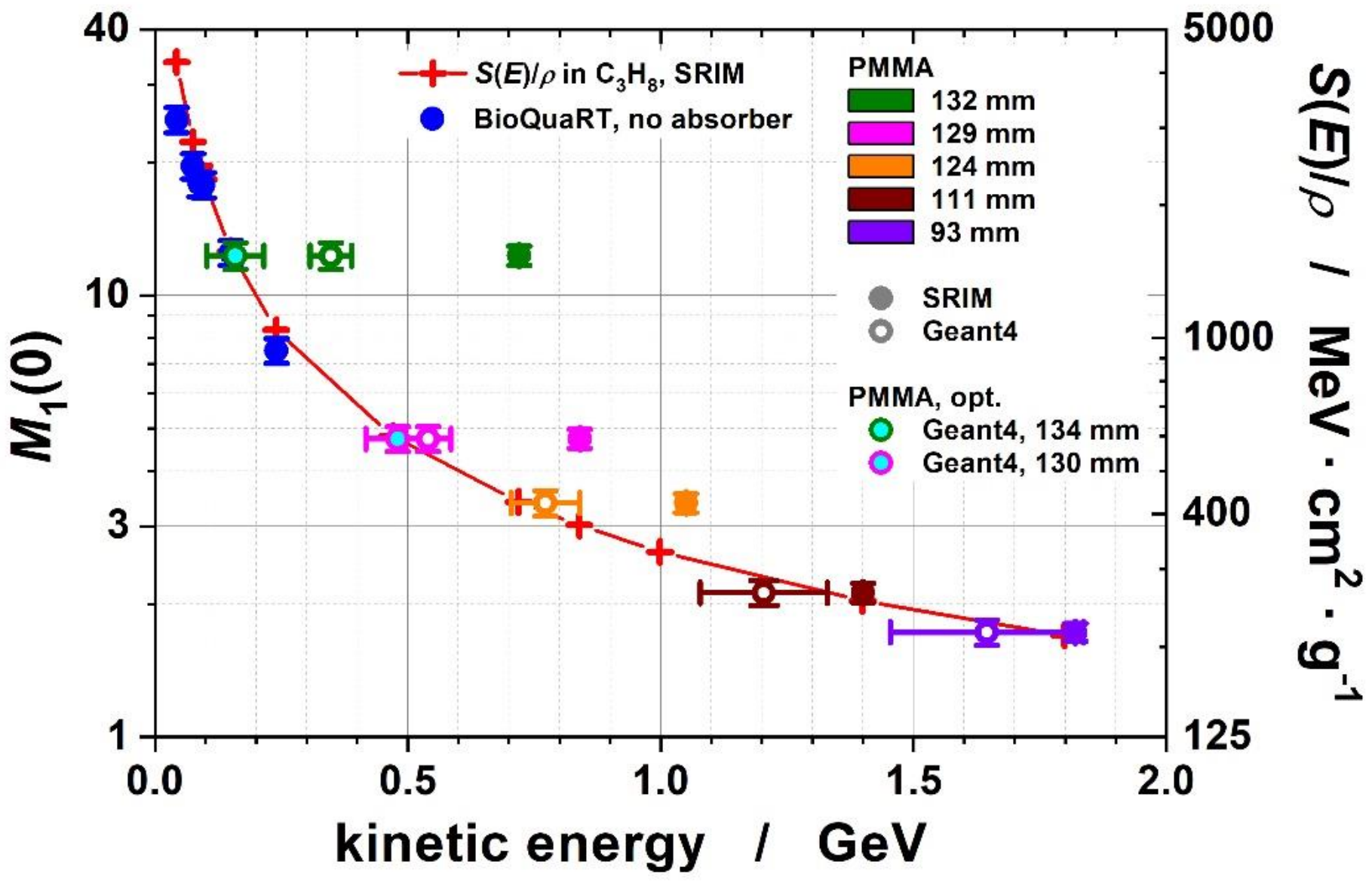

Figure 9: Energy dependence of the $M_1(0)$ values for the measurements with constant initial carbon ion energy and PMMA absorber of different thickness. The measured values are shown in filled circles as a function of the energy values calculated by SRIM and in open circles calculated by Geant4. Cyan-filled circles were used for energy values of additionally simulated absorber thicknesses, which best match the measurement results in the simulation. As reference data, earlier measurements from the BioQuaRT project [35] are shown in blue and mass stopping power $S(E)/\rho$ of $^{12}C$ ions in propane calculated with SRIM as a red line (right y-axis). (Adapted from [8])

### *4.2 Absorber thickness*

In Figure 7, the $M_1(0)$ values plotted as a function of absorber thickness of experiment and track structure simulation seemed to differ by a shift along the horizontal axis. Similarly, when comparing the measured ICSD for $d = 0$ with the simulated ICSD in Figure 8, a shift of the measured ICSD to higher ICS can be observed for high absorber thicknesses, particularly in the 132 mm case. This shift is reduced when the simulations are performed for an increased absorber thickness compared to the thickness used in the experiment. The required change of absorber thickness, which leads to an overlap of the distributions, amounts to 1 mm for the 124 mm and 129 mm cases, and 2 mm for the 132 mm case. Moreover, the shape of the simulated ICSDs leads to better agreement with the measured ones for slightly thicker absorbers for 129 mm and 132 mm absorber thickness (Figure 8(e) and (f)). When comparing the $M_1(0)$ values in Figure 7, assuming thicker absorbers also results in a reduced deviation between simulation and experiment. For smaller absorber thicknesses, the variation with thickness is shallow.

Tolerances in the manufacturing of PMMA blocks result in slight variations of the dimensions across the block, preventing it from achieving the same thickness at every point of its cross-section. However, these inaccuracies are in the sub-millimeter range and cannot account for deviations in the millimeter range. Furthermore, posterior mechanical measurements of the PMMA blocks employed in the experiments revealed that the combinations of blocks used to realize the designed absorbers had a total thickness slightly smaller than the targeted values (by a few tens of millimeters). Another possible explanation for the shift of the Bragg peak position to higher depths in the simulation could be a higher density of the PMMA used in the experiment compared to the simulated density of 1.18 g/cm³. PMMA is a hydrophilic material and can absorb up to 2 % water, leading to slightly different material properties [41]. Alternatively, a systematic offset of the cross-section data used in the simulations can also cause a shift in the Bragg peak position. Given that the discrepancy in the Bragg peak position is only 1.5 % of the absorber thickness, such an offset would only have to be of the same magnitude, which is small compared to the uncertainties that are generally assumed for nuclear interaction cross-sections [10].

### *4.3 Background events*

Heavy charged secondary particles contribute only marginally to the signal, as shown by an investigation of the mean stopping power of the most frequent heavy charged particles in the TVA in Figure 5. This analysis distinguishes between events where a carbon ion hits the second PSD and triggers signal acquisition and those where the carbon ion misses the second PSD. In triggered events, almost only carbon ions cross the TVA, while in non-triggered events many different particles can be found in the TVA. As presented in Table 1, signal triggering occurs in only 0.34 % to 1.75 % of events, with the proportion increasing as the absorber thickness decreases.

The high proportion of non-triggered events leads to a background signal due to particles that interact within the nanodosimeter during the measurement time window of a triggering carbon ion. This background was estimated based on track structure simulations of the particle interactions in the target volume of the nanodosimeter, as detailed in section 3.4. Two contributions to this background are distinguished: events where a carbon ion interacts in the target volume but misses the trigger detector (outlier events), and those where the carbon ion is absorbed by components upstream of the nanodosimeter, with only secondary particles interacting in the target volume (no-carbon event). The calculated background in $M_1$ ranges from $0.05 \pm 0.01$ to $0.30 \pm 0.08$, with the background signal increasing with absorber thickness. The largest proportion of the background is caused by events where the carbon ion interacts within the nanodosimeter (Table 2 and Table 3).

In the first paper [8], the measured $M_1(d)$ distributions were compared with previous measurements without absorbers conducted as part of the BioQuaRT project [35]. It was observed that the $M_1(d)$ values from the BioQuaRT measurements exhibit a steeper decrease with $d$ compared to the values from the current experiment. In the BioQuaRT measurements, the accelerated carbon ions only passed a thin gold foil in an evacuated scattering chamber (used for signal reduction via Rutherford scattering and beam monitoring) and the entrance window of the gas-filled nanodosimeter. Therefore, there was no contamination of the beam by other heavy charged particles, and the carbon ions entering the nanodosimeter were monoenergetic and well collimated due to a collimating aperture. Owing to the lower particle energy, the collimator was fully absorbing carbon ions entering its body and the projection of the collimator aperture from the secondary carbon ion source at the gold foil was underfilling the (20 x 3) mm² cross-sectional area of the trigger detector, which completely covered the spatial distribution of the carbon ions. Therefore, there was only a negligible background in the BioQuaRT data in contrast to the current experiment, thus leading to lower $M_1(d)$ values.

This absent background also caused a steeper decrease of $M_1(d)$ with $d$. This is demonstrated by comparing the simulated $M_1(d)$ values without background correction in Figure 10(a) for the current experiment and a measurement from the BioQuaRT project giving the same $M_1(0)$ as the present measurement with 132 mm absorber. Assuming an increase in the absorber thickness of 2 mm brings the simulated and BioQuaRT distribution close to agreement.

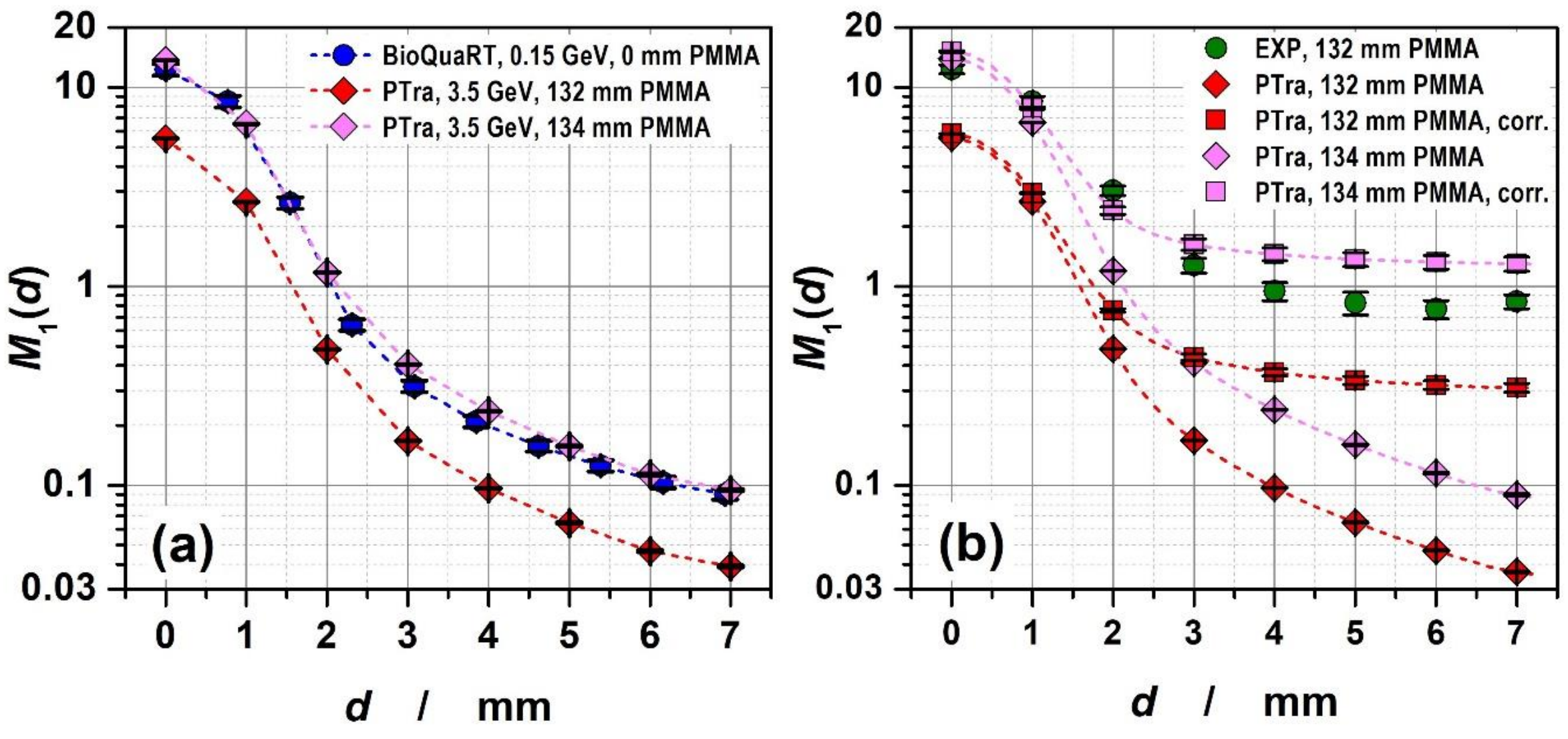

Figure 10. (a): Comparison of the results for $M_1(d)$ from the PTra simulations for 132 mm and 134 mm PMMA absorber thickness without background correction with a measurement from the BioQuaRT project [35] without PMMA absorber at lower carbon ion energy which produces the same $M_1(0)$ as the measurement shown on the left-hand side. (b): Comparison of the results for $M_1(d)$ from the PTra simulations for 132 mm and 134 mm PMMA absorber thickness with and without background correction with the measurements from the first paper [8] for the 132 mm absorber.

Assuming a larger effective absorber thickness also reduces the discrepancies between measured and simulated $M_1(d)$ distribution for high absorber thicknesses. This is shown in Figure 10(b) for a comparison of the measured $M_1(d)$ with 132 mm absorber thickness and simulations with absorber thicknesses of 132 mm and 134 mm.

### *4.4 Residual deviations between experiment and simulation*

The residual discrepancies between the measured $M_1(d)$ and those obtained after background correction from the simulations using the PMMA absorber thickness best reproducing the ICSDs are shown in Figure 11. Figure 11(a) shows the differences between the two data, and Figure 11(b) the ratio of these differences to the measured $M_1(d)$ values. For the case of the 132 mm absorber, the simulations for 134 mm thickness are slightly overestimating the $M_1$, as already seen in Figure 8(f) and Figure 10(b). Therefore, a large negative deviation can be seen in Figure 11(a) and (b) at impact parameters exceeding 2 mm. However, from Figure 8(f), it can be expected, that using a thickness somewhat smaller than 134 mm would have produced even better agreement with the measured data for 132 mm absorber thickness. This would have resulted in a reduction of the discrepancies at impact parameters exceeding 2 mm but would have presumably increased the differences and resulted in relative deviations at smaller impact parameters comparable to the other cases.

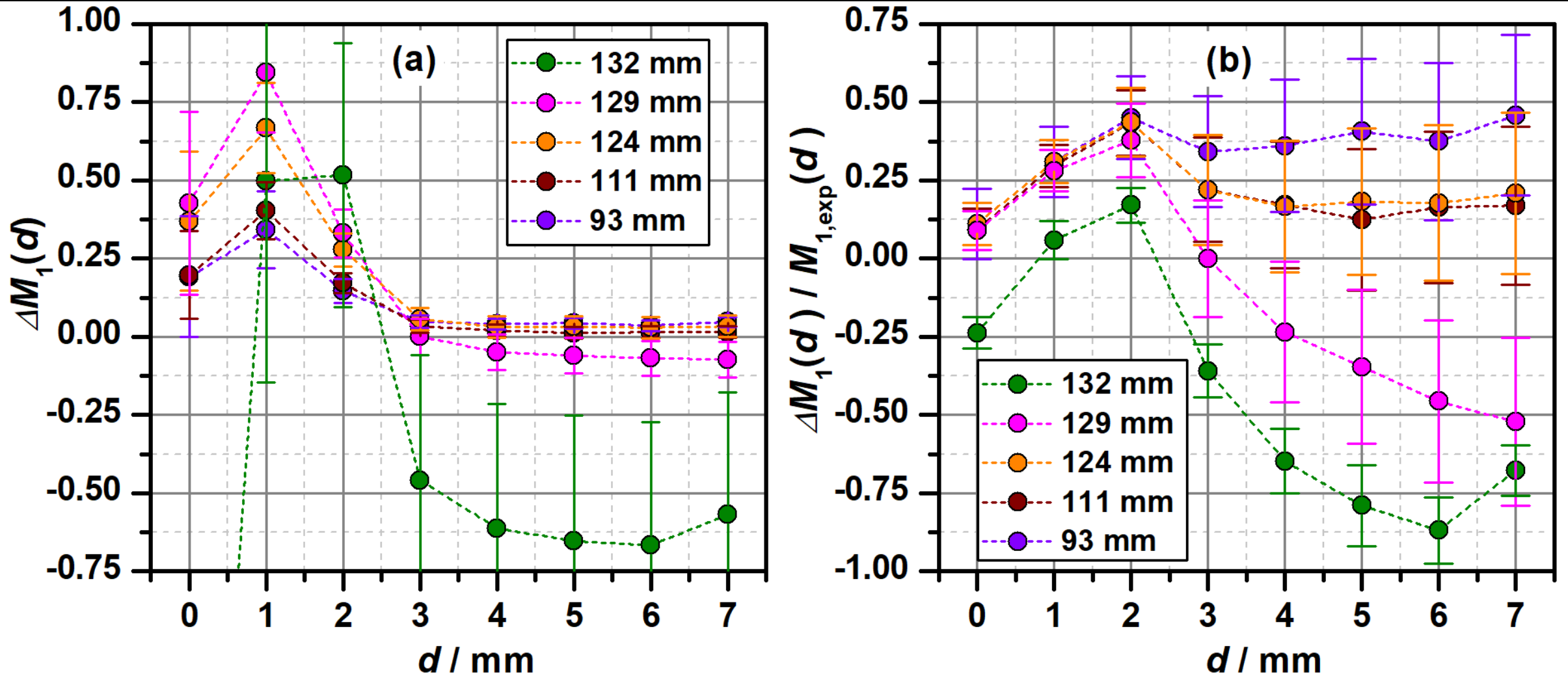

Figure 11: Residual differences between experiment and simulation: (a) absolute differences between the experimental $M_1(d)$ for the thickness of the PMMA absorber as given in the legend and the background-corrected simulated $M_1(d)$ for the PMMA thickness giving the best fit with the experimental ICSDs in Figure 8, namely 125 mm instead of 124 mm, 130 mm instead of 129 mm, and 134 mm instead of 132 mm; (b) the corresponding relative deviations.

Figure 11(b) shows up to 50 % deviation of experiment and simulation at the larger impact parameters, where Figure 6 showed the signal to be dominated by the background originating in coincidences between triggered and non-triggered events. As the background estimate was rather crude, it may be over or underestimated by several tens of percent. However, even if underestimated by a factor of two, the background would be still too small to account for the discrepancies seen in Figure 11(a) and (b) at impact parameters up to 2 mm.

Therefore, particle tracks contributing to these deviations must be correlated with the carbon ion triggering the measurement and belong to the same event. However, all secondary charged particles were considered in the PTra simulations.

It is remarkable that the main differences can be seen in Figure 11(a) for impact parameters below 3 mm, namely when the carbon ion trajectory traverses the target volume. This suggests that the differences may be related to the interaction probability of the carbon ion in the target volume. This interaction probability could be higher in the experiments than in the simulations due to three reasons.

The first reason would be a different density of propane molecules in the target volume if the gas pressure in the interaction volume was higher than the set value or the temperature was lower. However, to explain a difference in the measured $M_1$ values by more than 10 %, the pressure, temperature or both would have to be off by the same order of magnitude. For temperature, this can be ruled out as the temperature of the gas is essentially the same as in the laboratory. For pressure, this would require the calibration factor of the pressure gauge to have changed by several 10 %. This appears unlikely and contradicts the observed repeatability of the measurements, which was within ±6 % [8].

A second reason could be that the interaction cross-sections used in the PTra code are too low. As in many other codes, such as Geant4 [42], only ionizing interactions are considered in PTra for ions heavier than helium. Cross-sections for charge-transfer are expected to be negligibly small for carbon ions moving at several tens of percent of the speed of light, much higher than the kinetic energies of electrons in the molecular orbitals. However, electronic excitations are also neglected. While their cross-sections are generally smaller than for ionizations [43], some electronic excitations may decay by autoionization producing an additional ion at the position where the carbon ion interacted.

The total ionization cross-section for propane ($C_3H_8$) is determined in PTra using the model of Rudd et al. [44], scaling their parameter values for methane ($CH_4$) by the ratio of the number of valence electrons and by an additional empirical factor of 1.16 [43]. This factor was necessary to match the electron-impact ionization cross-sections of methane and propane in the several keV-range. This additional factor indicates that the contribution of K-shell ionizations may be more important for the total ionization cross-section at higher energies. Vacancies in the K-shell of low-Z atoms are predominantly (about 95 % [45]) filled by Auger-Meitner transitions that produce an additional electron and a doubly charged ion. This ion will either capture an electron from a neutral gas molecule during a collision or dissociate into two charged fragments. Both processes result in two ions instead of one and increase the mean ICS. Furthermore, the partial ionization cross-sections for the different molecular orbitals are adopted from those for electron-impact collisions [43]. As electrons are indistinguishable, the electron collisions depend on an exchange potential absent when ions collide with the same target. Therefore, the relative contributions of the different molecular shells may differ between ion and electron impact. This does not change the number of secondary ions produced in collisions of carbon ions but may change the number of ions produced by secondary electrons as their energy depends on the ionized shell.

The third possible reason for the discrepancy related to the passage of the carbon ion through the target volume is a systematic underestimation of the extraction efficiency. This appears unlikely based on previous investigations in which measurements and simulations were found to agree when the efficiency map obtained with the software package SIMION [46] was used to score secondary ions in the simulations [34,47,35,48–51]. One difference between the present and preceding simulations was that the radial dependence of the efficiency values at the discrete grid was linearly interpolated, not as a function of the radius but rather as a function of the square of the radius. In this way, the Gaussian shape of the extraction efficiency near the axis of the extraction aperture is better captured. This changed approach leads to a small increase in the scored number of ionizations and therefore cannot be responsible for the smaller $M_1$ values found in the simulations. On the other hand, the fact that the relative discrepancy between experiment and simulation increases with impact parameter could indicate a potential issue with the extraction efficiency map.

It should be noted that the track structure simulations used the particle energy spectrum from the Geant4 simulations. This means that the discrepancy between experiment and simulation is independent of the SRIM calculations and not explained by the possible causes of the incorrect energy values of the SRIM calculation outlined above.

Another factor influencing the simulation results may be the simplified geometry. Both simulations neglect the metal walls of the vacuum chamber, which is justified since they are far from the target region and mostly shielded by the capacitor plates. The capacitor plates are close to the target region and interactions of particles hitting the capacitor plates producing secondary electrons may result in additional ionizations in the interaction volume that are not considered in the simulations. However, to be effective and contribute to the signal, these ionizations must occur in the target volume. Therefore, only secondary electrons produced by charged particles impinging on the lower electrode near the extraction aperture could potentially lead to extra ions produced in the target volume by these secondary electrons. The trajectories of heavy charged particles were screened for impact on the capacitor plate and the points of impact were found to be uniformly distributed across the plates. Hits occurred within a 5 mm radius around the extraction aperture only for about $10^{-4}$ of the no-carbon events, below $10^{-5}$ of the outlier events and none of the triggered events. Therefore, this additional background contribution that was present in the experiments and omitted in the simulations can be assumed to be negligibly small compared to the components shown in (Table 2 and Table 3).

On the other hand, some bias could be caused by the two-step simulation approach. The starting point of the trajectories at the entrance of the nanodosimeter volume for the PTra simulations is determined by a linear projection based on the position and direction of the particles in the TVA calculated by Geant4. This method does not take any deviation from a straight-line trajectory by scattering into account. However, the probability of scattering is low, and thus small position uncertainties can be assumed. In addition, the track structure simulation starts the particles in the plane of the rear of PSD1 using as starting energy the energy that the particles had at the start of the condensed-history step crossing the TVA. The energy loss of the particles between PSD1 and the TVA is in the order of 10 keV (Figure 5), and thus this approximation is not expected to introduce a bias. Nevertheless, it would be interesting to repeat the simulations in a one-step approach when the cross-section data for propane will have been implemented in the Geant4-DNA extension.

### *4.5 Concluding thoughts about the nanodosimetric measurement concept.*

The simulation results show (Supplementary Figure S3) that, as intended, gas ion collection in the experiments was only triggered when a carbon ion hit the second PSD. Triggering by other heavy charged particles was prevented by applying a window discriminator. This may appear somewhat at variance with the concept of nanodosimetry where all charged particles producing ionization clusters would have to be considered. Using a hypothetical large trigger detector that allowed all heavy charged particles to be unambiguously detected, one could conceive an experiment determining the contributions of the different tracks separately.

However, the carbon ions and most of their secondary particles have kinetic energies of 10 % or more of their rest energy. Therefore, their speed is in the order of several 10 % of the speed of light. Consequently, the particles from the same event (primary carbon ion) traverse the interaction volume in less than 10 ns. Compared to the drift times of the secondary gas ions, which are in the order of 100 µs, their arrival is practically simultaneous, meaning that the contributions of the different particles in an event cannot be distinguished. The measured ICSD would simply be the average over all triggered events where those triggered by particles other than carbon ions may have a large background due to carbon ions missing the detector. Alternatively, if the trigger detector was sufficiently large to register all heavy charged particles in the subtended solid angle, it would not be possible to uniquely define an impact parameter since each particle would have its own. This implies that a nanodosimetric measurement in which all particles of the event were registered would not only face the challenge of realizing this registration but also require the concepts of nanodosimetry to be reconsidered. It is hoped that this may be possible along the lines of the theoretical approach recently suggested by Faddegon et al. [7].

## 5. Conclusion

In this paper, simulations have been presented to explain the unexpected results of nanodosimetric measurements with carbon ions, as presented in the first part of the paper [8].

It appears that simulations using SRIM may be insufficient to predict the energy loss of ions passing through thicker layers of material, such as the PMMA absorber employed in the measurements to mimic different depths in a phantom.

Furthermore, it was found that the (2x10) mm² trigger detector only covered a small portion of the spatial distribution of the carbon ions in the detector. The high proportion of untriggered carbon ions results in a background signal when an untriggered carbon ion interacts in coincidence with a triggered carbon ion in the nanodosimeter. This background leads to increased $M_1$ values and a different $d$-dependence of the $M_1(d)$ distribution. For future experiments, a large-area trigger detector should be used to cover the entire spatial distribution.

The quantitative comparison between the measurements and track structure simulations indicate that the assumption of a higher absorber thickness leads to a better agreement between experiment and simulation for thick absorbers. Uncertainties in the cross-section data for carbon interactions in PMMA in the order of 1 % are assumed to be the most likely cause. A bias of 10 % to 20 % in the cross-section for the interaction of carbon ions in propane at the energies encountered in the experiment could also account for the residual small underestimation of the measured results by the simulations. However, it could not be ruled out that this underestimation may have resulted from the two-step simulation approach. This warrants further investigation and emphasizes the need for the implementation of the cross-section data for propane in Geant4-DNA to enable track-structure simulation in propane-based detectors in a single simulation.

## 6. Acknowledgments

The authors thank the dedicated team of the High-Performance Computing Cluster at PTB for their ongoing support throughout the Geant4 simulations. Markus Bär is accredited for providing access to the compute cluster operated by his department and Oliver Hensel for support with technical issues. This work was partly supported by the "Metrology for Artificial Intelligence in Medicine (M4AIM)" program, funded by the German Federal Ministry of Economic Affairs and Climate Action in the frame of the QI-Digital Initiative.

## Supplement

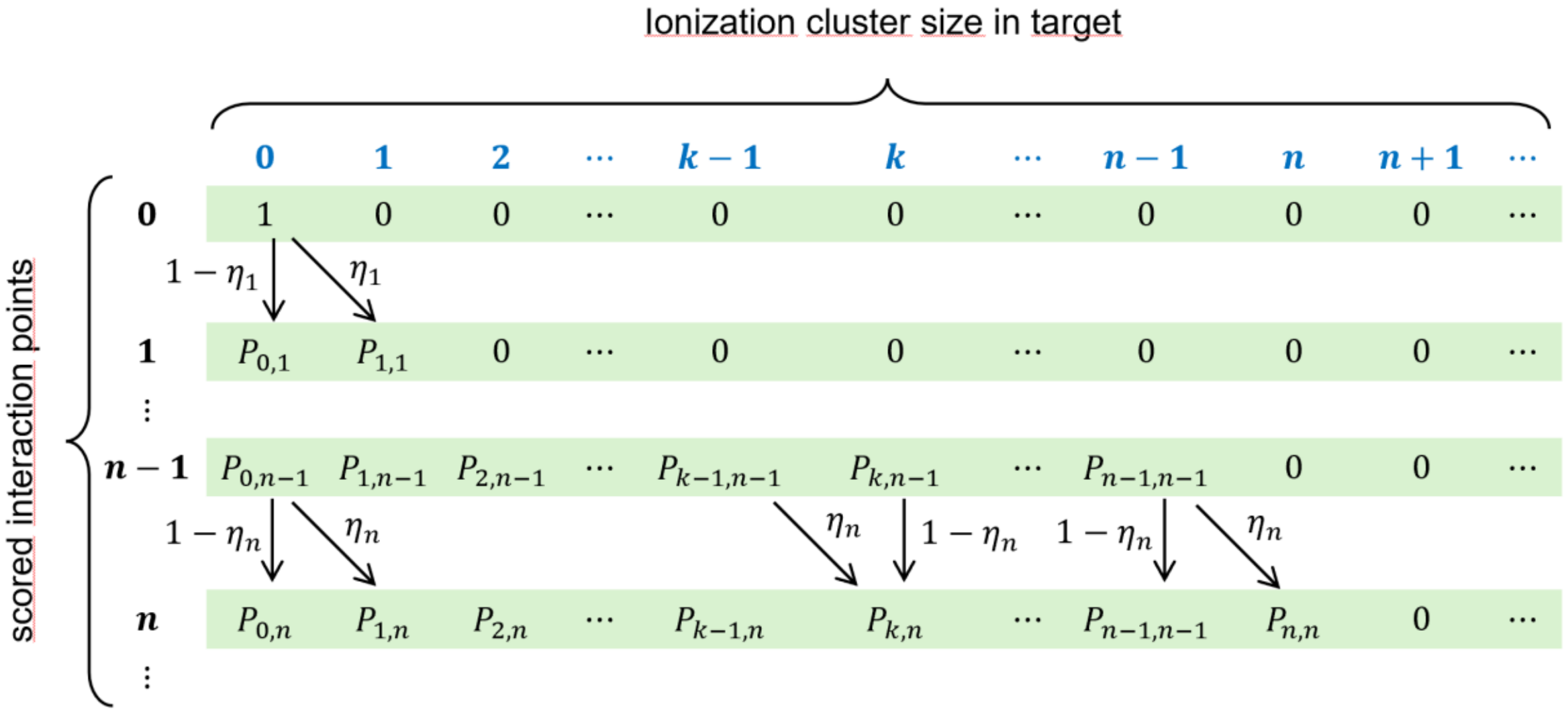

Supplementary Figure S1: Illustration of the approach for direct construction of the ionization cluster size distribution (ICSD) in a target from a single track. The highlighted rows are the ICSD for the track after processing the number of energy transfer points (ETPs) shown in bold left of the row. The column headers in bold indicate the ICS and the entries in the highlighted cells indicate the corresponding probability. Initially, the ICSD has a probability of 1 for ICS 0 and zero for all other values of ICS. For each energy transfer point (ETP) within the range of the efficiency map, the probability of the ion being detected is obtained from its position relative to the origin of the efficiency map. The respective value $\eta_n$ for the $n$-th ETP is used to update the ICSD based on the values of the preceding step (row).

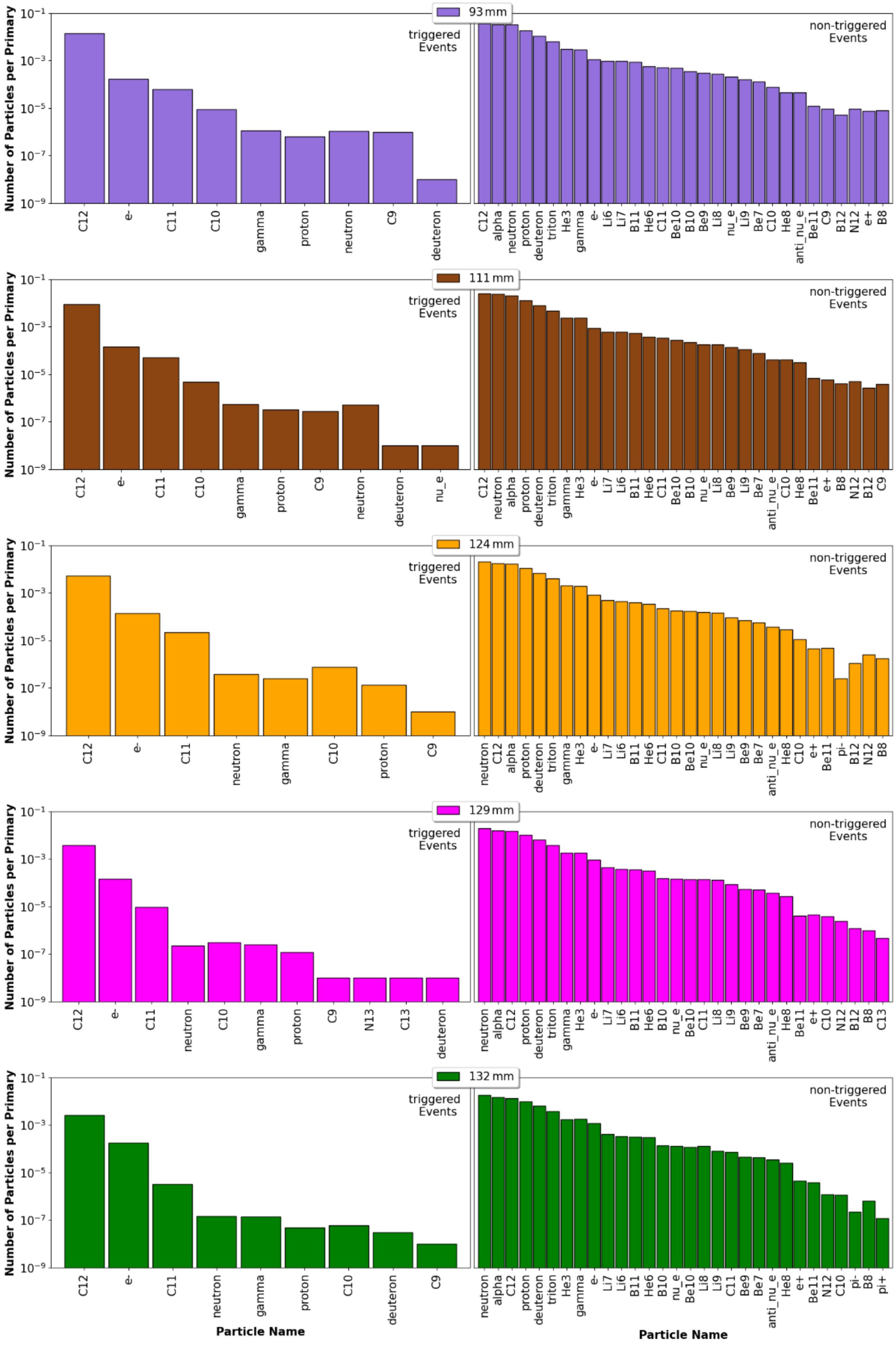

Supplementary Figure S2: Frequency per primary carbon ion of the most abundant secondary particles passing the target volume aperture for the different absorber thicknesses used in the experiment. Left plots: triggered events; right plots: non-triggered events.

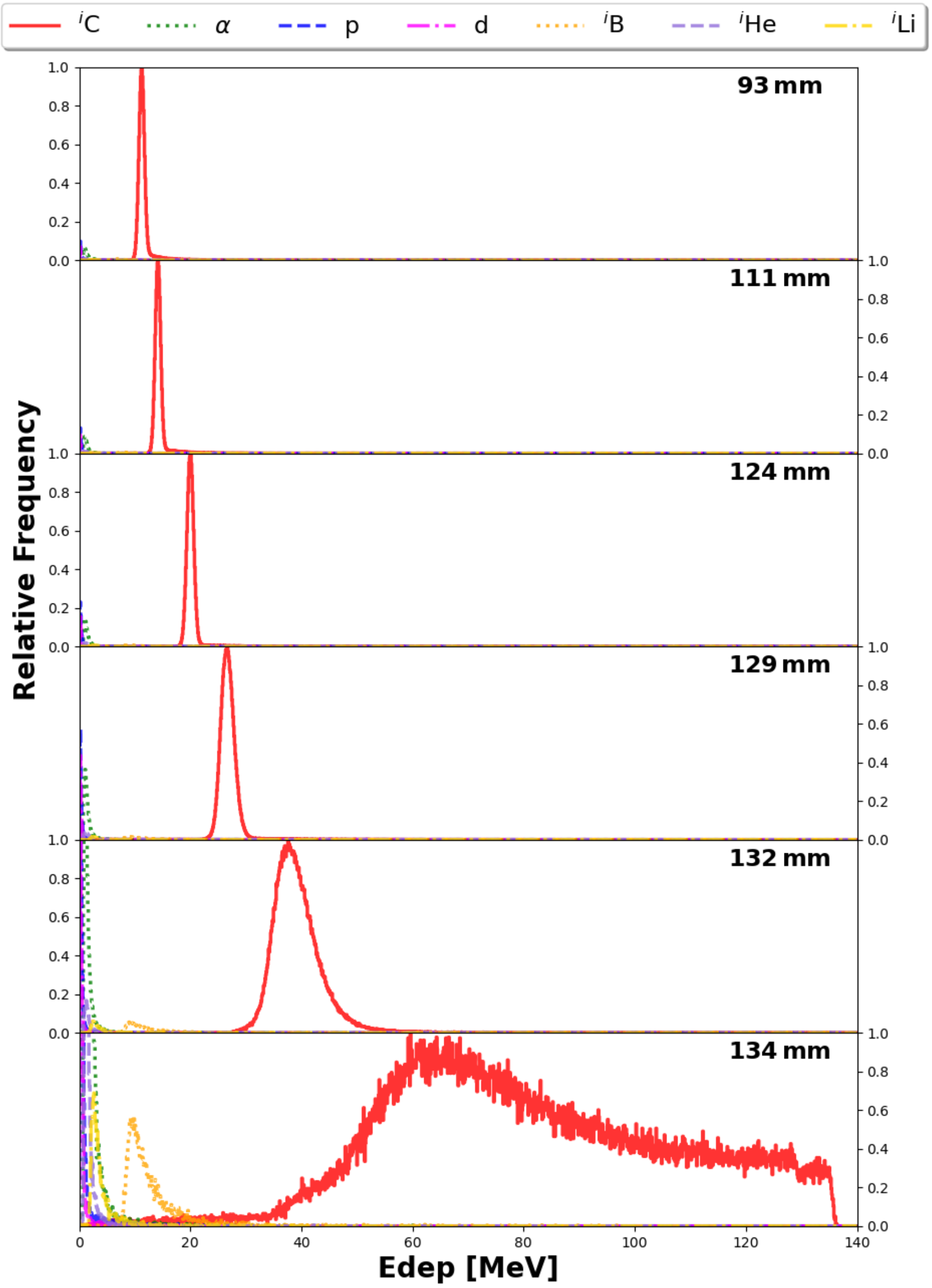

Supplementary Figure S3: Energy deposition per event of the most frequent particles in the second PSD (trigger detector) for the different absorber thicknesses used in the experiment and the absorber thickness determined in the simulations, which best reproduces the measured data of the 132 mm case.

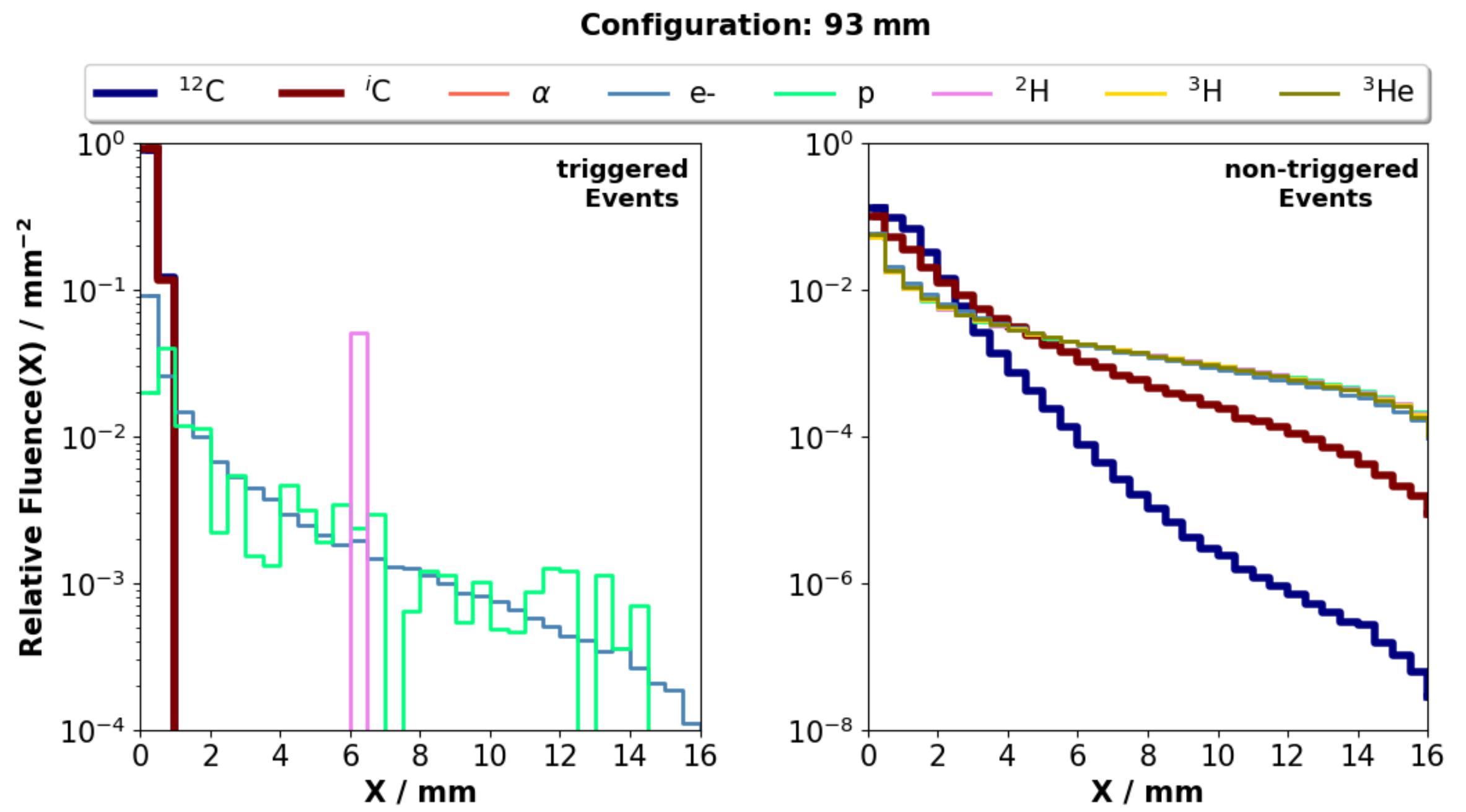

Supplementary Figure S4: Radial distribution of the fluence through a circular ring with an outer and inner radius corresponding to the respective upper and lower bin edges of the carbon ions and most frequent secondary particles crossing the target volume aperture (TVA) for an absorber thickness of 93 mm. The histograms are normalized to the number of respective particles traversing the TVA. Left: events with a carbon ion hitting PSD2; right: events without hit in PSD2.

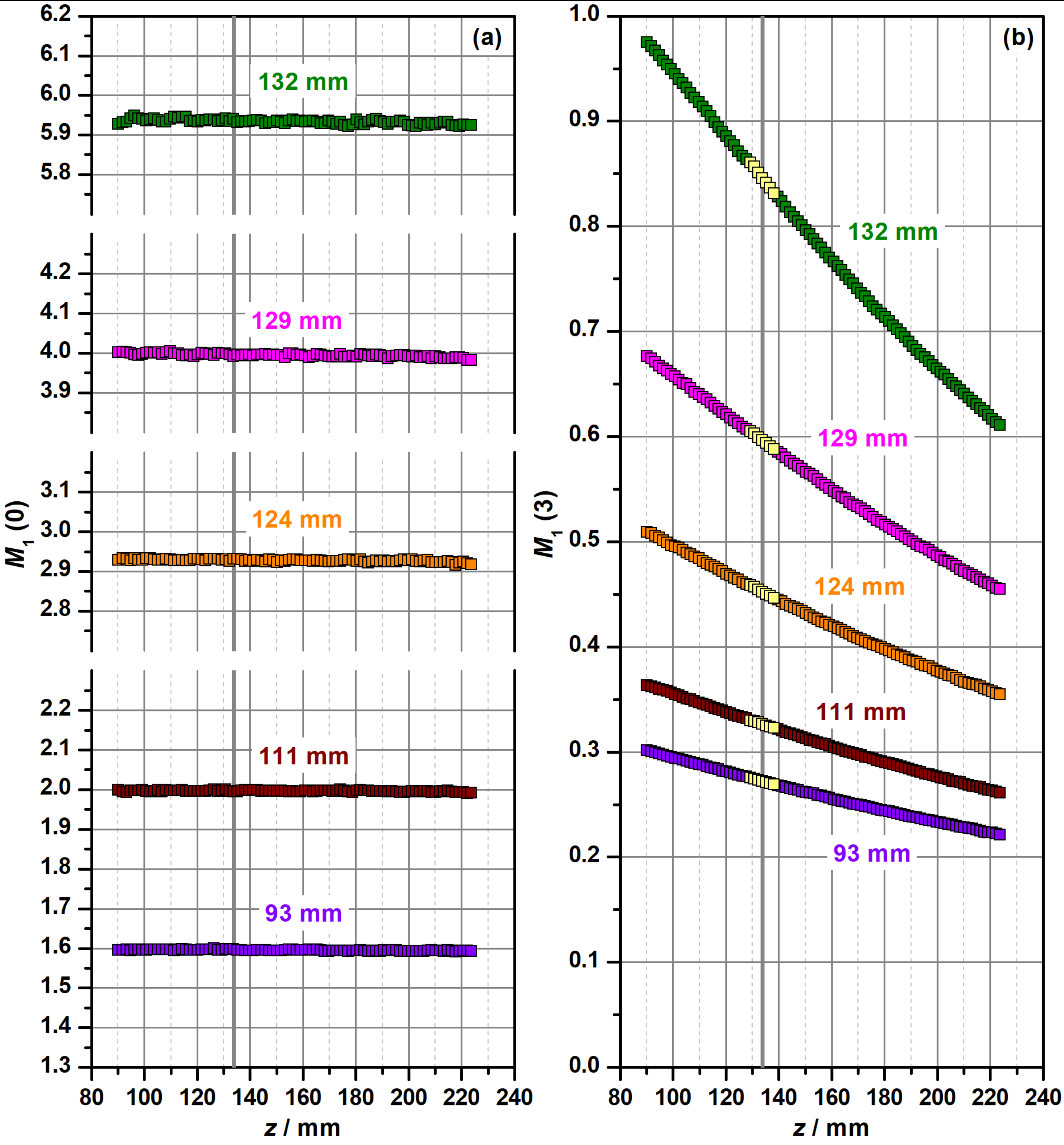

Supplementary Figure S5: (a) Variation of the mean ionization cluster size $M_1$ at impact parameter $d = 0$ mm found in the track structure simulations with PTra when considering only triggered events, i.e. events with a carbon ion traversing the second silicon detector. (b) Variation of $M_1$ at 3 mm offset from the center of the PSDs for outlier events, i.e. events with a carbon ion missing the trigger detector. This offset corresponds to the location of the extraction aperture in the experiments. The thick vertical gray line indicates the location of the extraction aperture in the experiments. The symbols in yellow indicate the data points considered for determining the background estimate.